\documentclass[]{aastex631}

\usepackage{hyperref}
\usepackage{graphicx}
\usepackage{float}
\usepackage[caption=false]{subfig}
\usepackage{enumitem}
\usepackage{comment}

\newcommand{\rev}[1]{{\textcolor{black}{#1}}}

\shorttitle{Lifting the Hazy Veil of GJ 1214 b}
\shortauthors{Ohno, Schlawin, Bell et al.}

\begin{document}

\title{A Possible Metal-Dominated Atmosphere Below the Thick Aerosols of GJ 1214 b Suggested by its JWST Panchromatic Transmission Spectrum}

\correspondingauthor{Kazumasa Ohno}
\email{ohno.k.ab.715 AT gmail.com}

\author[0000-0003-3290-6758]{Kazumasa Ohno}
\affil{Division of Science, National Astronomical Observatory of Japan, 2-12-1 Osawa, Mitaka-shi 1818588 Tokyo, Japan}

\author[0000-0001-8291-6490]{Everett Schlawin}
\affiliation{Steward Observatory, 933 North Cherry Avenue, Tucson, AZ 85721, USA}

\author[0000-0003-4177-2149]{Taylor J. Bell}
\affiliation{Bay Area Environmental Research Institute, NASA's Ames Research Center, Moffett Field, CA 94035, USA}
\affiliation{Space Science and Astrobiology Division, NASA's Ames Research Center, Moffett Field, CA 94035, USA}

\author[0000-0002-8517-8857]{Matthew M. Murphy}
\affiliation{Steward Observatory, 933 North Cherry Avenue, Tucson, AZ 85721, USA}

\author[0000-0002-9539-4203]{Thomas G. Beatty}
\affiliation{Department of Astronomy, University of Wisconsin--Madison, Madison, WI 53703, USA}

\author[0000-0003-0156-4564]{Luis Welbanks}
\affiliation{School of Earth and Space Exploration, Arizona State University, Tempe, AZ, USA}

\author[0000-0002-8963-8056]{Thomas P. Greene}
\affiliation{Space Science and Astrobiology Division, NASA's Ames Research Center, Moffett Field, CA 94035, USA}

\author[0000-0002-9843-4354]{Jonathan J. Fortney}
\affiliation{Department of Astronomy and Astrophysics, University of California, Santa Cruz, CA, USA}

\author[0000-0001-9521-6258]{Vivien Parmentier}
\affiliation{Laboratoire Lagrange, Observatoire de la Côte d’Azur, CNRS, Université Côte d’Azur, Nice, France}

\author[0000-0001-8745-2613]{Isaac R. Edelman}
\affiliation{Bay Area Environmental Research Institute, NASA's Ames Research Center, Moffett Field, CA 94035, USA}





\author[0000-0001-6086-4175]{Nishil Mehta}
\affiliation{Université Côte d'Azur, Observatoire de la Côte d'Azur, CNRS, Laboratoire Lagrange, France}


\author[0000-0002-7893-6170]{Marcia J. Rieke}
\affiliation{Steward Observatory, 933 North Cherry Avenue, Tucson, AZ 85721, USA}



\begin{abstract}
GJ1214b is the archetype sub-Neptune for which thick aerosols have prevented us from constraining its atmospheric properties for over a decade.
In this study, we leverage the panchromatic transmission spectrum of GJ1214b established by HST and JWST to investigate its atmospheric properties using a suite of atmospheric radiative transfer, photochemistry, and aerosol microphysical models.
We find that the combined HST, JWST/NIRSpec and JWST/MIRI spectrum can be well-explained by atmospheric models with an extremely high metallicity of [M/H]$\sim3.5$ and an extremely high haze production rate of $F_{\rm haze}{\sim}10^{-8}$--${10}^{-7}~{\rm g~cm^{-2}~s^{-1}}$.
Such high atmospheric metallicity is suggested by the relatively strong CO$_2$ feature compared to the haze absorption feature or the CH$_4$ feature in the NIRSpec-G395H bandpass of $2.5$--$5~{\rm {\mu}m}$.
The flat $5$--$12~{\rm {\mu}m}$ MIRI spectrum also suggests a small scale height with a high atmospheric metallicity that is needed to suppress a prominent $\sim6~{\rm {\mu}m}$ haze feature.
We tested the sensitivity of our interpretation to various assumptions for uncertain haze properties, such as optical constants and production rate, and all models tested here consistently suggest extremely high metallicity.
Thus, we conclude that GJ1214b likely has a metal-dominated atmosphere where hydrogen is no longer the main atmospheric constituent.
We also find that different assumptions for the haze production rate lead to distinct inferences for the atmospheric C/O ratio.
We stress the importance of high precision follow-up observations to confirm the metal-dominated atmosphere, as it challenges the conventional understanding of interior structure and evolution of sub-Neptunes.

\end{abstract}


\section{Introduction} \label{sec:intro}

Sub-Neptunes are the most common type of planet known in the Milky Way \citep{howard2012occurrenceKepler,fressin2013occurenceKepler,fulton2017radiusGap,dattilo2023unifiedTreatmentKeplerOccurence}.
It is thus crucial to investigate them to establish a unified understanding of planet formation and evolution.
However, their nature remains elusive. 
Even their bulk compositions have been uncertain since various kinds of interior structure, such as a rocky interior with a thick hydrogen-helium envelope and an icy core with a thin steam envelope, can explain the same planetary mass-radius relation \citep[e.g.,][]{adams2008OceanVsThickAtm,rogers2010threeOriginsGJ1214b,valencia2013bulkCompositionGJ1214b}.

Atmospheric observations, such as transmission spectroscopy, provide clues to the chemical makeup of exoplanets. 
Many previous studies observed transmission spectra of sub-Neptunes using the Hubble Space Telescope (HST) over the last decade. 
However, these observations struggled to constrain the atmospheric properties of sub-Neptunes, since many of them typically exhibit very weak spectral features of atmospheric molecules \citep[e.g.,][]{kreidberg14,knutson2014gj436,benneke2019gj3470LowMetallicityMieScattering,libby-roberts19_superpuff,Dymont+22,Brande+24}. 
These featureless spectra can be attributed to the presence of optically thick clouds and/or hazes at high altitudes that prevent us from probing the atmosphere below these aerosol layers.

The James Webb Space Telescope (JWST) has opened a novel window to access the chemical compositions of exoplanetary atmospheres, including several sub-Neptunes \citep[for review, see][]{kempton2024transitingAtmospheresJWST}. 
JWST has the advantage of observing the spectrum at long wavelengths with unprecedented precision, where clouds and hazes are relatively transparent. 
Indeed, JWST transmission spectroscopy has revealed that some sub-Neptunes, namely K2-18b, TOI-270d, and GJ3470b, have atmospheric metallicities ${\sim}100\times$ the solar value \citep{madhusudhan2023k2d18,benneke2024toi270dMiscibleMetalRich,holmberg2024toi270d,beatty24gj3470b}.
The inferred metal-rich atmospheres are roughly consistent with planet formation models that predicted metal enrichment due to planetesimal ablation in protoatmospheres \citep[e.g.,][]{fortney2013lowMassP,Venturini16_enrichment}.
Meanwhile, the aforementioned planets have relatively clear atmospheres. 
JWST still observed featureless spectra in some sub-Neptunes, including TOI-836c and LHS1140b, which could be attributed to thick aerosol layers and/or extremely metal-rich atmospheres \citep{wallack2024jwstCompass,Cadieux24_LHS1140,Damiano+24_LHS1140}.
The atmospheric compositions of such heavily cloudy/hazy sub-Neptunes remain largely uncertain.

GJ1214b is the canonical sub-Neptune discovered by the MEarth project \citep{charbonneau2009gj1214bDiscovery}. 
GJ1214b has a relatively low bulk density ($2.2~{\rm g/cm^3}$, $R_{\rm p}=2.73R_{\rm \oplus}$ and $M_{\rm p}=8.17M_{\rm \oplus}$, \citealt{cloutier2021gj1214bPreciseMass}) that is consistent with a rocky core surrounded by a hydrogen-rich envelope or an icy core surrounded by a metal-rich envelope \citep{rogers2010threeOriginsGJ1214b,valencia2013bulkCompositionGJ1214b,Nixon+24internalStructureGJ1214bJWST}.
Thanks to the small radius of its host star and its nearby location (14.6 pc), in theory, GJ1214b serves as an ideal target to characterize sub-Neptune atmospheres.
However, intense observational efforts have revealed that GJ1214b shows a remarkably flat spectrum that is attributed to thick aerosol layers in the upper atmosphere \citep[e.g.,][]{bean10,desert2011_gj1214b_spitzer,Berta+2012_gj1214b_flat,Fraine2013_Spitzer_GJ1214b,kreidberg14}.
The flat spectrum of GJ1214b motivated many modeling studies on the formation of mineral clouds \citep{Morley2013_gj1214b,morley2015superEclouds,Charnay+15GJ1214bspectrum,Ohno&Okuzumi18microphysicalCloudModelsGJ214gj436b,gao2018microphysicsGJ1214clouds,Ohno+20fluffyaggregate,Christie2022_gj1214b_gcm} and photochemical hazes \citep{Morley2013_gj1214b,morley2015superEclouds,Kawashima&Ikoma18,Kawashima&Ikoma19,Adams+19aggregateHazes,Lavvas+19photochemicalHazes}.
Most of those studies suggested that the atmosphere of GJ1214b presumably has a metallicity ${\ga}100\times$ solar to explain a sufficiently flat spectrum.

Recent JWST observations provide new insight into the atmospheric properties of GJ1214b.
\citet{kempton2023reflectiveMetalRichGJ1214} observed a thermal phase curve of GJ1214b using JWST/MIRI-LRS and found that the planet shows a notably large amplitude of the phase curve. 
Previous studies used global circulation models to show that the phase curve amplitude increases with increasing metallicity \citep{Kataria+14gcmGJ1214b,Charnay+15GJ1214bspectrum,zhang2017bulkCompDynamics}.
The large amplitude of GJ1214b thus strongly indicates high atmospheric metallicity. \citet{kempton2023reflectiveMetalRichGJ1214} also observed the transmission spectrum and emission spectra from dayside/nightside with JWST/MIRI-LRS. 
While \citet{kempton2023reflectiveMetalRichGJ1214} reported the tentative detection of H$_2$O from the dayside/nightside emission spectra, they found that the MIRI transmission spectrum is still consistent with a flat line.
\citet{gao2023gj1214} analyzed the JWST/MIRI transmission spectrum combined with the HST/WFC3 transmission spectrum using a microphysical model of photochemical hazes and suggested a high atmospheric metallicity of $\ga1000\times$ solar and an extremely high haze production rate of $\ga{10}^{-10}~{\rm g~cm^{-2}~s^{-1}}$.
Most recently, \citet{schlawin2024gj1214b_placeholder} used JWST/NIRSpec-G395H to observe the transmission spectrum of GJ1214b and reported a moderate detection of CO$_2$ and CH$_4$.
They discussed that the relative strengths of CO$_2$ and CH$_4$ features likely indicate a very high atmospheric metallicity.

In this study, we leverage the panchromatic transmission spectrum of GJ1214b established by HST/WFC3, JWST/NIRSpec-G395H, and JWST/MIRI-LRS to further investigate its atmospheric properties. 
This paper is a companion study of \citet{schlawin2024gj1214b_placeholder}. 
We utilize a suite of atmospheric radiative transfer, photochemistry, and microphysical models to constrain atmospheric compositions and aerosol properties in GJ1214b.
The organization of this paper is as follows.
In Section \ref{sec:method_model} and Appendix \ref{sec:appendix_method1}, we introduce the atmospheric models that are used to interpret the panchromatic spectrum.
In Section \ref{sec:results}, we explain the results of the model--data comparison and their interpretation.
We test a number of different model assumptions to enhance the robustness of the inferred atmospheric properties.
We also discuss the implications of inferred atmospheric properties for planet formation and evolution processes.
In Section \ref{sec:summary}, we summarize our findings.


\section{Method Overview}\label{sec:method_model}
We interpret the transmission spectrum of GJ1214b observed by HST/WFC3 \citep{kreidberg14}, JWST/MIRI-LRS \citep{kempton2023reflectiveMetalRichGJ1214}, and JWST/NIRSpec-G395H \citep{schlawin2024gj1214b_placeholder}.
To this end, we construct a forward model grid using radiative transfer, photochemisty, and aerosol microphysical models under the assumption that photochemical haze is the predominant aerosol in GJ1214b, as in previous studies \citep{Adams+19aggregateHazes,Lavvas+19photochemicalHazes,gao2023gj1214}. 
We call this grid a photochemical-microphysical (PM) haze grid and mainly use it to interpret the panchromatic transmission spectrum.

For the PM haze grid, we describe the modeling details in Appendix \ref{sec:appendix_method1} and overview the modeling approach here.
We first use a radiative-convective equilibrium model called \texttt{EGP} code used in previous studies \citep[e.g.,][]{Marley&McKay99thermalStructure,Fortney+05comparitivePlanetaryAtmospheres,Morley2013_gj1214b,marley2015rev,gao2020aerosolsSilicatesAndHazes,OhnoFortney2021_Nitrogen1} to calculate the atmospheric temperature-pressure (TP) profile for various atmospheric metalicities [M/H] \footnote{We refer [M/H] to describe [(C+O)/H], a base-10 log of carbon+oxygen-to-hydrogen ratio normalized by solar value, unless otherwise indicated.}. 
For each TP profile, we use a publicly available photochemical model \texttt{VULCAN} \citep{Tsai+17,tsai2021VULCANcomparitiveStudy} to calculate the vertical distributions of molecular abundances, where we vary the eddy diffusion coefficient and atmospheric C/O ratio as additional free parameters. 
For haze properties, we use a two-moment microphysical model developed by \citet{Ohno&Okuzumi18microphysicalCloudModelsGJ214gj436b} and \citet{Ohno+20fluffyaggregate} to calculate the vertical distributions of the mean particle size and the mass mixing ratio of haze particles for various eddy diffusion coefficients and the column-integrated haze production rate, as done in \citet{Ohno&Kawashima20superRaleighSlopes}.
In this study, we assume compact spherical haze particles for simplicity.
Note that TP profiles in this study do not reflect radiative feedback from photochemistry and haze formation, since it makes a vast parameter survey challenging.
We leave the self-consistent treatment of TP profiles, photochemistry, and haze formation to future studies.
Based on the PM haze grid, we use a publicly available radiative transfer model \texttt{CHIMERA} \citep{line2013chimera} to calculate the transmission spectrum to compare with the observed spectrum.
We use the public Mie theory code \texttt{PyMieScatt} \citep{Sumlin+18} to calculate haze optical properties with refractive indices of the haze analogue produced from $1000\times$ solar metallicity gas at 400 K \citep[water-rich tholin,][]{He+23organichazesWaterRich}. 
Note that the experimental temperature of $400~{\rm K}$ is consistent with the temperature of the upper atmosphere inferred by \citet{kempton2023reflectiveMetalRichGJ1214} for the GJ1214b's dayside, where haze formation likely takes place.
We test different refractive indices of exoplanetary haze analogs in Section \ref{sec:many_optic}.
For model-data comparisons, we perform grid-based retrieval using the \texttt{PyMultinest} \citep{feroz2009multinest,Buchner+14_pymultinest} to obtain the posterior distributions of atmospheric metallicity [M/H], C/O ratio, column-integrated haze production rate $F_{\rm haze}$ and eddy diffusion coefficient at $1~{\rm bar}$ $K_{\rm zz,1bar}$.
\rev{A few previous studies attempted to estimate the haze production rate $F_{\rm haze}$ from photochemical model \citep{Kawashima&Ikoma19,Lavvas+19photochemicalHazes}.
Although we test the similar approach in Section \ref{discussion:Fhaze}, our fiducial analysis agnostically varies $F_{\rm haze}$ given the large uncertainty in the formation process of exoplanetary hazes.
Our analysis in Section \ref{discussion:Fhaze} and Appendix \ref{Appendix:photolysis} also suggests that OCS plays a critical role in producing haze particles at high metallicity atmospheres, whereas the OCS recombination rate is highly uncertain for the temperature relevant to GJ1214b \citep{tsai2021VULCANcomparitiveStudy}.
}
We refer the reader to Appendix \ref{sec:appendix_method1} for more details of the atmospheric model and fitting procedure.

\section{Results}\label{sec:results}

\subsection{Overview of Forward Model Results}

\begin{figure}
    \centering
    \includegraphics[width=\linewidth]{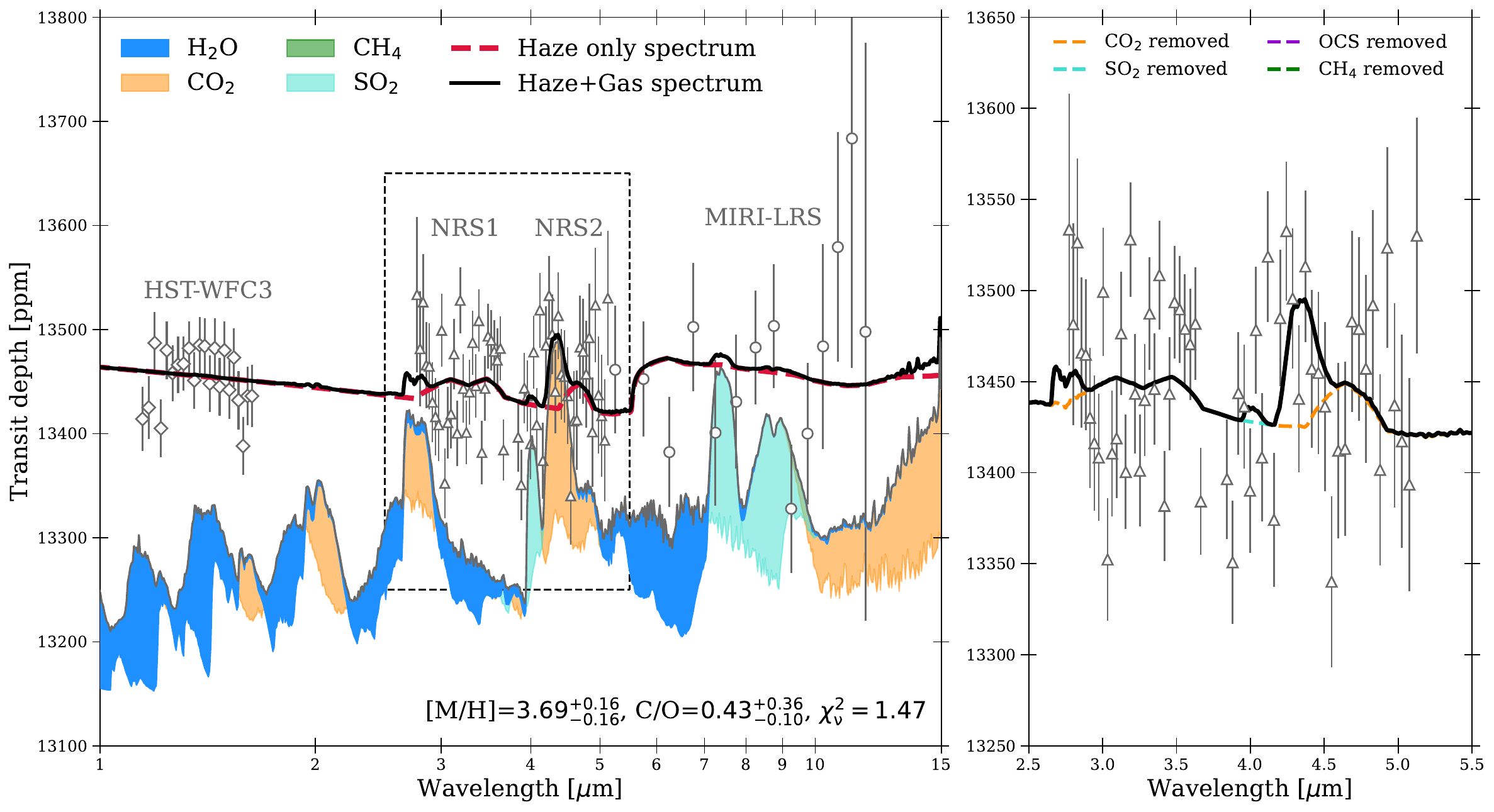}
    \caption{(Left) A median model of the PM haze grid retrieval on the panchromatic GJ1214b spectrum. The black solid line shows the spectrum that includes both haze and gas opacity. The gray solid and red dashed lines show the spectra including only gas and haze opacity, respectively, to illustrate their individual contributions. The color shaded regions denote the differences in the haze-free spectrum induced if we remove the opacity of a certain molecule, highlighting which molecule can contribute the atmospheric spectrum. The PM haze grid obtains median and $1\sigma$ uncertainty values of [M/H]$=3.69^{+0.16}_{-0.16}$ and C/O$=0.43^{+0.36}_{-0.10}$, where the reduced chi-squared value is $\chi^2_{\rm \nu}=1.47$ for 88 degrees of freedom (96 datapoints, 8 model parameters). Note that each observed spectrum has been shifted according to the retrieved instrumental offset parameters. (Right) The same median spectrum zoomed in the NIRSpec/G395H band. The colored dashed lines show the spectrum that removes the opacity of certain molecular species.}
    \label{fig:median_haze_He400}
\end{figure}

Figure \ref{fig:median_haze_He400} shows the median transmission spectrum from the PM haze grid.
Although the atmospheric model produces H$_2$O, CO$_2$, and SO$_2$ features, most of them are completely masked by haze opacity.
For instance, H$_2$O absorption is expected to contribute to the spectrum at HST band and $\sim5$--$7~{\rm {\mu}m}$, but the haze obscures the H$_2$O features and makes the spectrum completely featureless. 
Because the haze controls the overall shape of the observed spectrum, the model spectrum shows absorption features of haze particles themselves at $\sim3$--$3.6~{\rm {\mu}m}$, $\sim4.6~{\rm {\mu}m}$, and $\sim6~{\rm {\mu}m}$ \citep[for function groups causing these fetures, see][]{He+23organichazesWaterRich}.
The $\sim6~{\rm {\mu}m}$ feature originating from C=O and C=N stretching \citep{He+23organichazesWaterRich} is particularly prominent, although its presence is unclear in the current MIRI spectrum.

Although haze mostly controls the overall shape of observed spectrum, some molecular features still contribute to the NIRSpec spectrum at $2.5$--$5~{\rm {\mu}m}$.
The most notable is the CO$_2$ feature at $4.3~{\rm {\mu}m}$, which is consistent with \citet{schlawin2024gj1214b_placeholder} who reported the highest detection significance for CO$_2$ among all of the molecules considered based on the NIRSpec/G395H transmission spectrum.
CO$_2$ also moderately affects the spectrum at $\sim2.7~{\rm {\mu}m}$, which is not inconsistent with the current data.
\rev{The model-data comparison suggests an extremely high metallicity of [M/H]$=3.69^{+0.16}_{-0.16}$, which is consistent with \citet{Hu&Seager14} who predicted that abundant CO$_2$ is possible only at hydrogen-depleted atmospheres for GJ1214b conditions.}
The inferred high metallicity with low C/O ratio promotes the formation of SO$_2$ \citep{Tsai2023_SO2,Crossfield23,beatty24gj3470b} that slightly affects the spectrum at $\sim4~{\rm {\mu}m}$ and $\sim7.5~{\rm {\mu}m}$, although its contribution is too small to be realized in the current data.

\citet{schlawin2024gj1214b_placeholder} reported $2\sigma$ detection of CH$_4$ from the structured NIRSpec spectrum at $\sim3.3~{\rm {\mu}m}$. 
However, within the PM haze grid, the absorption feature of organic hazes could produce a similar structure at $\sim3$--$3.6~{\rm {\mu}m}$ without CH$_4$ when the atmosphere is extremely hazy. 
The current NIRSpec data struggle to distinguish between the absorption features of haze and CH$_4$.
We note that haze produces another feature at $\sim4.6~{\rm {\mu}m}$, which resembles the current NIRSpec data around this wavelength.

Our model-data comparison suggests an extremely high atmospheric metallicity on GJ1214b.
Our grid-based retrieval using the entire panchromatic data set obtains the atmospheric metallicity of ${\rm [M/H]}=3.69^{+0.16}_{-0.16}$, the carbon-to-oxygen ratio of ${\rm C/O}=0.43^{+0.36}_{-0.1}$, the haze mass flux of $\log_{\rm 10}(F_{\rm haze}{\rm [g~cm^{-2}~s^{-1}]})=-7.61^{+0.41}_{-0.77}$, and the eddy diffusion coefficient of $\log_{\rm 10}(K_{\rm zz,1bar}{\rm [cm^{2}~s^{-1}]})=5.34^{+0.35}_{-0.23}$.
The high atmospheric metallicity obtained here supports the high metallicity suggested by a large amplitude of the MIRI phase curve \citep{kempton2023reflectiveMetalRichGJ1214} as well as the suggestions of previous modeling studies  for aerosol formation \citep{gao2018microphysicsGJ1214clouds,Lavvas+19photochemicalHazes,Ohno+20fluffyaggregate,gao2023gj1214}.
We elaborate why the current GJ1214b spectrum prefers the extremely high metallicity in the next Section.

\begin{figure}
    \centering
    \includegraphics[width=\linewidth]{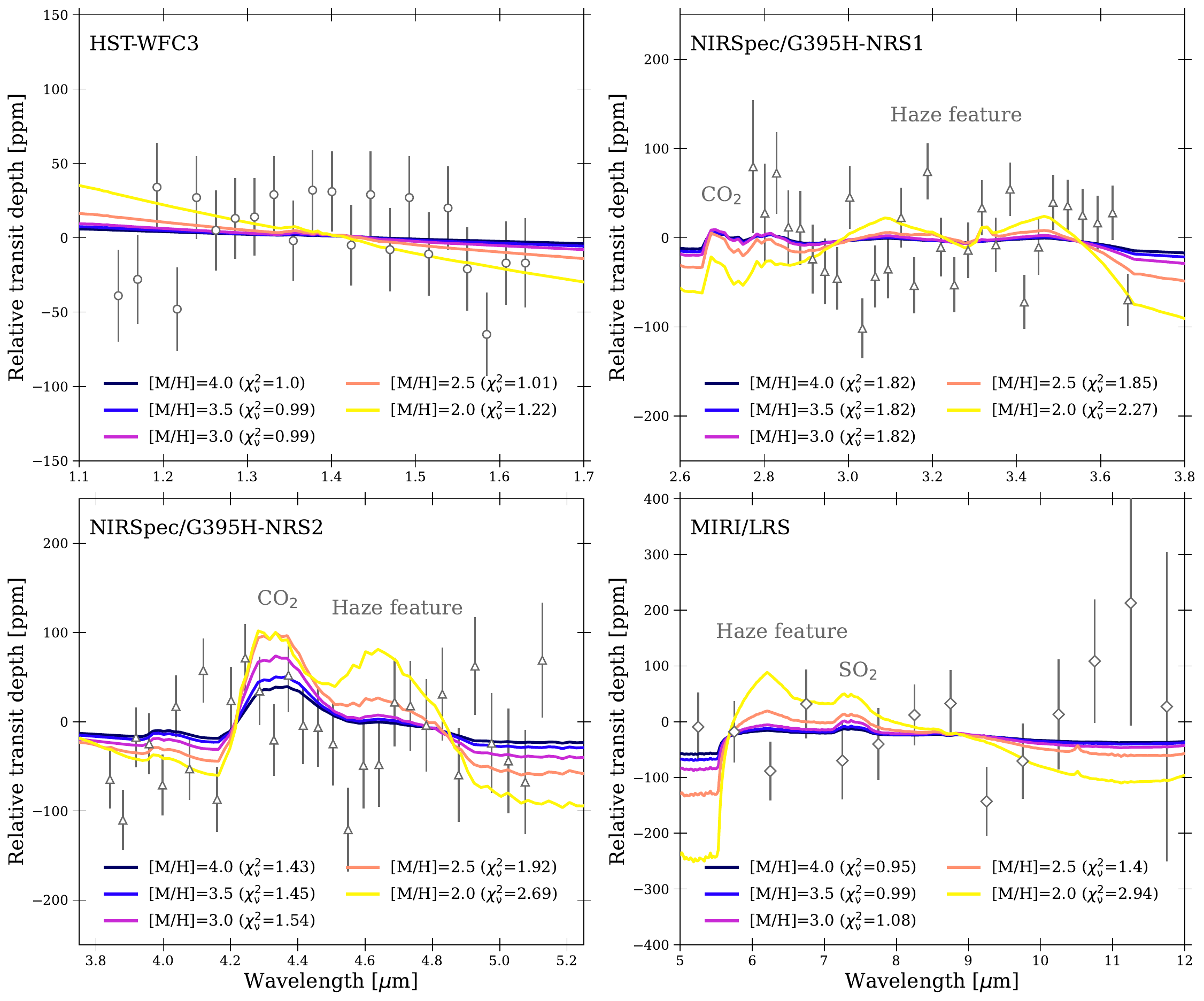}
    \caption{Instrument-level model-data comparisons for different atmospheric metallicities. The data points in each panel are the transit depths relative to their averaged value at each instrument band. Each colored lines show the best-fit model for atmospheric metallicity of [M/H]=2.0, 2.5, 3.0, 3.5, and 4.0, where C/O ratio, column-integrated haze production rate, and eddy diffusion coefficient are set to the values of the median model for panchromatic spectrum.}
    \label{fig:trans_metal}
\end{figure}

\subsection{Why does GJ1214b's Spectrum Prefer an Extreme Metallicity?}\label{sec:why_highZ}
To clarify why the GJ1214b spectrum prefers extremely high metallicity, we compared the atmospheric model with the relative transit depths within each HST, G395H/NRS1, G395H/NRS2, and MIRI band for different atmospheric metallicities.  
We perform the model-data comparison on relative transit depths within each instrumental band individually so that the unknown instrumental offsets do not bias the metallicity inference.
For each atmospheric metallicity, we first adjust the reference planetary radius at $P=10~{\rm bar}$ so that each metallicity model produces approximately the same transit depth at $4.3~{\rm {\mu}m}$ where CO$_2$ feature exists.
We then subtract a constant offset from the model spectrum to calculate the relative transit depths for each instrument band.
The spectum is then compared with the observed transit depths relative to their averaged depths at each instrumental band.
The constant offset is selected to minimize the chi-squared values within each instrumental band for each atmospheric metallicity.
The atmospheric C/O ratio, haze mass flux, and eddy diffusion coefficients are set to the values of the median model of the PM haze grid with water-rich tholin to isolate metallicity effects.

Figure \ref{fig:trans_metal} shows the model-data comparisons in each instrumental band for atmospheric metallicities of [M/H]=2.0, 2.5, 3.0, 3.5, and 4.0.
Given the extremely high haze mass flux, all metallicity models produce featureless spectra within the HST band. 
A spectral slope is steeper at lower metallicity owing to a lower mean molecular weight, and thus a larger scale height.
However, all models are nearly equally consistent with the observed HST spectrum, although the HST spectrum slightly prefers a higher atmospheric metallicity.

JWST spectra show more remarkable differences due to atmospheric metallicities.
In the NIRSpec/G395H-NRS1 band, the haze absorption feature is too large to be consistent with the data if the metallicity is as low as [M/H] = 2.0 due to the too large scale height.
NRS1 spectrum also shows that the depth at $\sim2.8~{\rm {\mu}m}$ is comparable to or greater than the depth at $\sim3.2~{\rm {\mu}m}$.
Although the error bar is large, this trend can be interpreted to mean that the $\sim2.8~{\rm {\mu}m}$ CO$_2$ feature is comparable to or greater than the haze feature.
Since the CO$_2$ abundance increases with increasing metallicity, a higher atmospheric metallicity tends to produce a better fit to the NRS1 data.
\rev{We note that \citet{Hu&Seager14} also predicted that CO$_2$ could be abundant compared to CH$_4$ only under hydrogen-depleted environments for GJ1214 b (see their Figure 5 and Section 4).}

The NRS2 spectrum also prefers a higher atmospheric metallicity for similar reasons.
In the NRS2 band, CO$_2$ produces a bump at $\sim4.3~{\rm {\mu}m}$ while haze produces a bump at $\sim4.6~{\rm {\mu}m}$ if the atmosphere is extremely hazy.
Although the data show only weak wavelength dependence of the transit depths at $>4.6~{\rm {\mu}m}$, the $\sim4.6~{\rm {\mu}m}$ haze feature becomes too large at [M/H]=2.0 and 2.5.
Atmospheric metallicity of [M/H]$\ge3.0$ allows the weak wavelength variation at $>4.6~{\rm {\mu}m}$.
Moreover, a higher metallicity also leads $\sim4.3~{\rm {\mu}m}$ CO$_2$ feature to be greater than the $\sim4.6~{\rm {\mu}m}$ haze feature, which is also in more line with the NRS2 spectrum.


The MIRI-LRS spectrum also strongly prefers the high atmospheric metallicity. 
This is because the MIRI spectrum is almost flat \citep{kempton2023reflectiveMetalRichGJ1214}, while an extremely hazy atmosphere is expected to produce a prominent organic feature around $6~{\rm {\mu}m}$ \citep{Corrales+23photochemHazesCtoO,He+23organichazesWaterRich}.
As a result, the lower atmospheric metallicity of [M/H]=2.0--2.5 leads to a too large haze feature to fit the flat MIRI spectrum due to the large scale height.
High metallicity models of [M/H]$\ga$3.0 can only result in a small enough haze feature to fit the MIRI spectrum.

\subsection{Influences of Haze Optical Constants}\label{sec:many_optic}
\begin{figure}
    \centering
    \includegraphics[width=\linewidth]{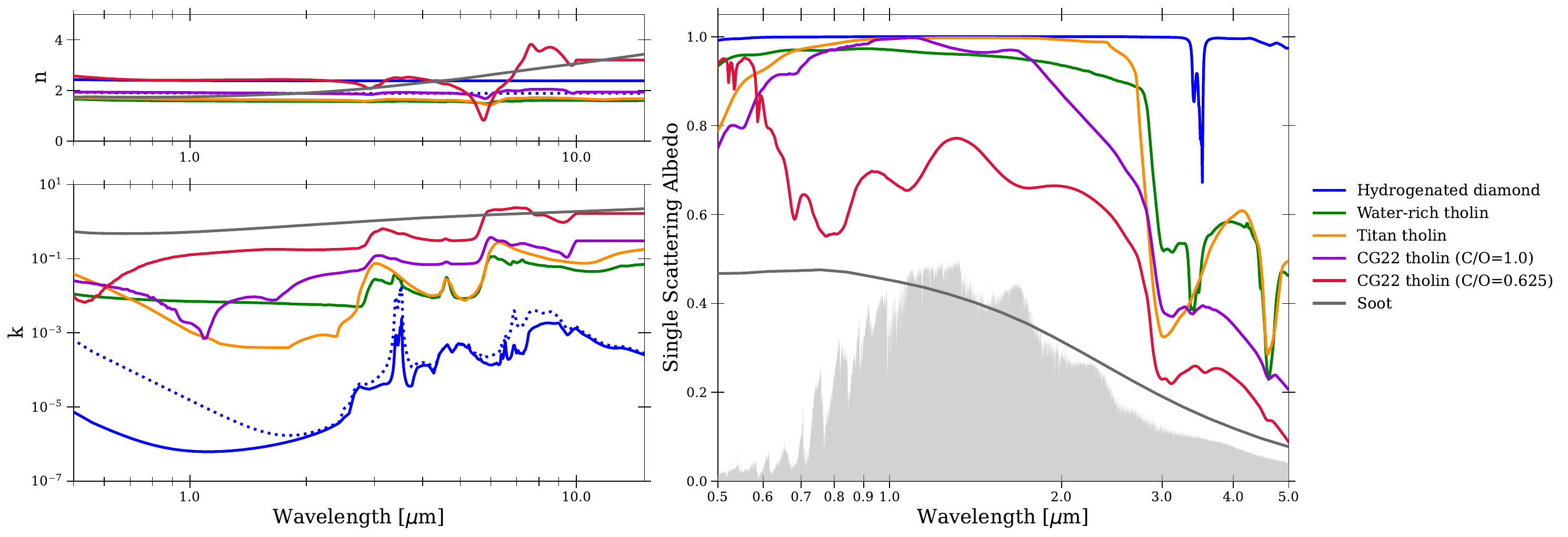}
    \includegraphics[width=\linewidth]{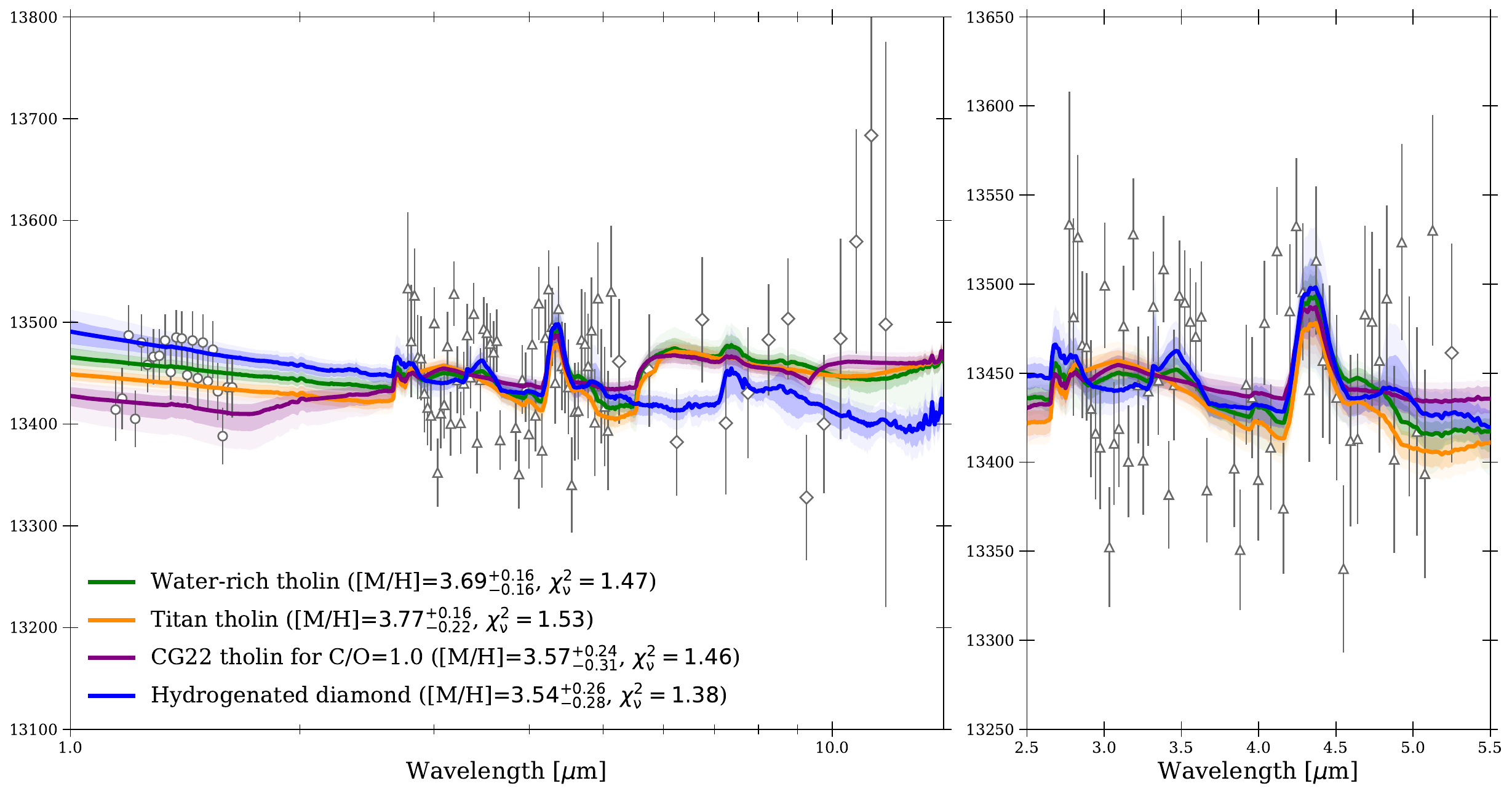}
    \caption{ (Left top) Real and imaginary parts of refractive indices, $n$ and $k$, for various substances that might represent exoplanetary hazes. The refractive indices are taken from hydrogenated diamond \citep{Jones&Ysard22_diamond_optics}, water-rich tholin at $400~{\rm K}$ \citep{He+23organichazesWaterRich}, Titan tholin \citep{Khare+84opticalConstantsTholins}, tholin produced from N$_2$+CH$_4$+CO$_2$ gas mixture (CG22 tholin) for C/O=1.0 and 0.625 \citep{Corrales+23photochemHazesCtoO}, and soot \citep{Lavvas&Koskinen17}. For hydrogenated diamonds, the solid and dotted lines show the refractive indices for the particle radius of $0.01$ and $0.1~{\rm {\mu}m}$, respectively, as the indices depend on the particle size that controls the fraction of the hydrogenated surface within bulk volume \citep{Jones&Ysard22_diamond_optics}. (Right top): Single scattering albedo of a particle with a radius of $0.3~{\rm {\mu}m}$ for each refractive index. The gray shaded region denotes the shape of the GJ1214 spectrum in terms of ${\lambda}F_{\rm \lambda}$, taken from the MUSCLE survey \citep{France+16_MUSCLE}, which diagnoses at which wavelength the incident stellar lights are mainly deposited. (Bottom) Median spectra of PM haze grid retrievals with various refractive indices of exoplanetary haze analogs. The shaded regions denote the $1\sigma$ and $2\sigma$ variations. The observed spectrum at each instrument has been shifted according to the offset parameters obtained by the grid retrieval with water-rich tholin. Within the refractive indices tested here, hydrogenated diamonds achieves the lowest value of $\chi^2_{\rm \nu}=1.38$.}
    \label{fig:trans_many_haze}
\end{figure}

The preceding argument may raise concerns that our interpretation strongly depends on the choice of haze refractive indices.
To assess the robustness of our interpretation against the haze refractive indices, we repeat the PM haze grid retrieval with other refractive indices.
The left top panels of Figure \ref{fig:trans_many_haze} show the real and imaginary parts of complex refractive indices of several exoplanetary haze analogs, which include Titan tholin \citep{Khare+84opticalConstantsTholins}, tholin deposited from N$_2$, CH$_4$, and CO$_2$ gas mixture for different C/O ratios \citep[CG22 tholin,][]{Corrales+23photochemHazesCtoO} and soot \citep{Lavvas&Koskinen17}.
We also add the refractive indices of hydrogenated diamond \citep{Jones&Ysard22_diamond_optics}, which was recently proposed as a potential candidate of ''reflective haze`` in close-in planets \citep{ohno2024diamond} owing to the similarities between exoplanetary atmospheres and the environments of low-pressure diamond synthesis through chemical vapor deposition \citep[e.g.,][]{Angus&Hayman88,Bachmann+91,Butler+09}.
From the MIRI phase curve, \citet{kempton2023reflectiveMetalRichGJ1214} suggested that aerosols in GJ1214b should have a single-scattering albedo of $\ga0.8$ over the wavelength range of GJ1214's peak luminosity.
As shown in the right top panel of Figure \ref{fig:trans_many_haze}, this requirement could be satisfied by hydrogenated diamond, water-rich tholin, Titan tholin and CG22 tholin for C/O=1.0 if we assume the haze particle radius of $0.3~{\rm {\mu}m}$ \footnote{This selection still applies if we assume the particle radius of $1~{\rm {\mu}m}$}.
Thus, we test those refractive indices to see how different refractive indices affect our metallicity inference. 

Consequently, we confirm that the metal-dominated atmosphere is robust to the choice of haze optical constants.
The bottom panels of Figure \ref{fig:trans_many_haze} show the median transmission spectrum of the PM haze grid retrieval for various optical constants of haze analogs. 
All of tested models suggest the atmospheric metallicity of [M/H]$\sim3.5$--$3.8$.
Each haze optical constant leads to a similar featureless spectrum at the HST band, while they show noticeable differences in the NIRSpec and MIRI bands.
For instance, Titan tholin produces a broad organic feature around $\sim3~{\rm {\mu}m}$, while the CG22 tholin produces an almost featureless spectrum in the same wavelength range.
Water-rich tholin, Titan tholin, and CG22 tholin all commonly produce a similar organic feature at $\sim6~{\rm {\mu}m}$.
On the other hand, hydrogenated diamonds lead to a quite different spectrum shape: they produce a narrow feature at $3.53~{\rm {\mu}m}$ due to C-H stretching on the hydrogenated surface \citep{Jones&Ysard22_diamond_optics} while it lacks the prominent $\sim6~{\rm {\mu}m}$ feature.
In general, different refractive indices lead to a different shape of the overall transmission spectrum; however, unknown offsets among each instrument prevent us from conclusively identifying which haze optical constant is most likely (see also Appendix \ref{Appendix:offset}).
Within the optical constants tested here, hydrogenated diamonds do the best job of fitting the current GJ1214b spectrum ($\chi^2_{\rm \nu}=1.38$) thanks to the $3.53~{\rm {\mu}m}$ feature, which is consistent with a bump seen around $\sim3.5~{\rm {\mu}m}$ in the NIRSpec spectrum, and the lack of prominent $\sim6~{\rm {\mu}m}$ feature, which is in line with the flat MIRI spectrum.

\subsection{\rev{Revisiting The Haze Mass Budget}}\label{discussion:Fhaze}
\begin{figure}
    \centering
    \includegraphics[width=0.95\linewidth]{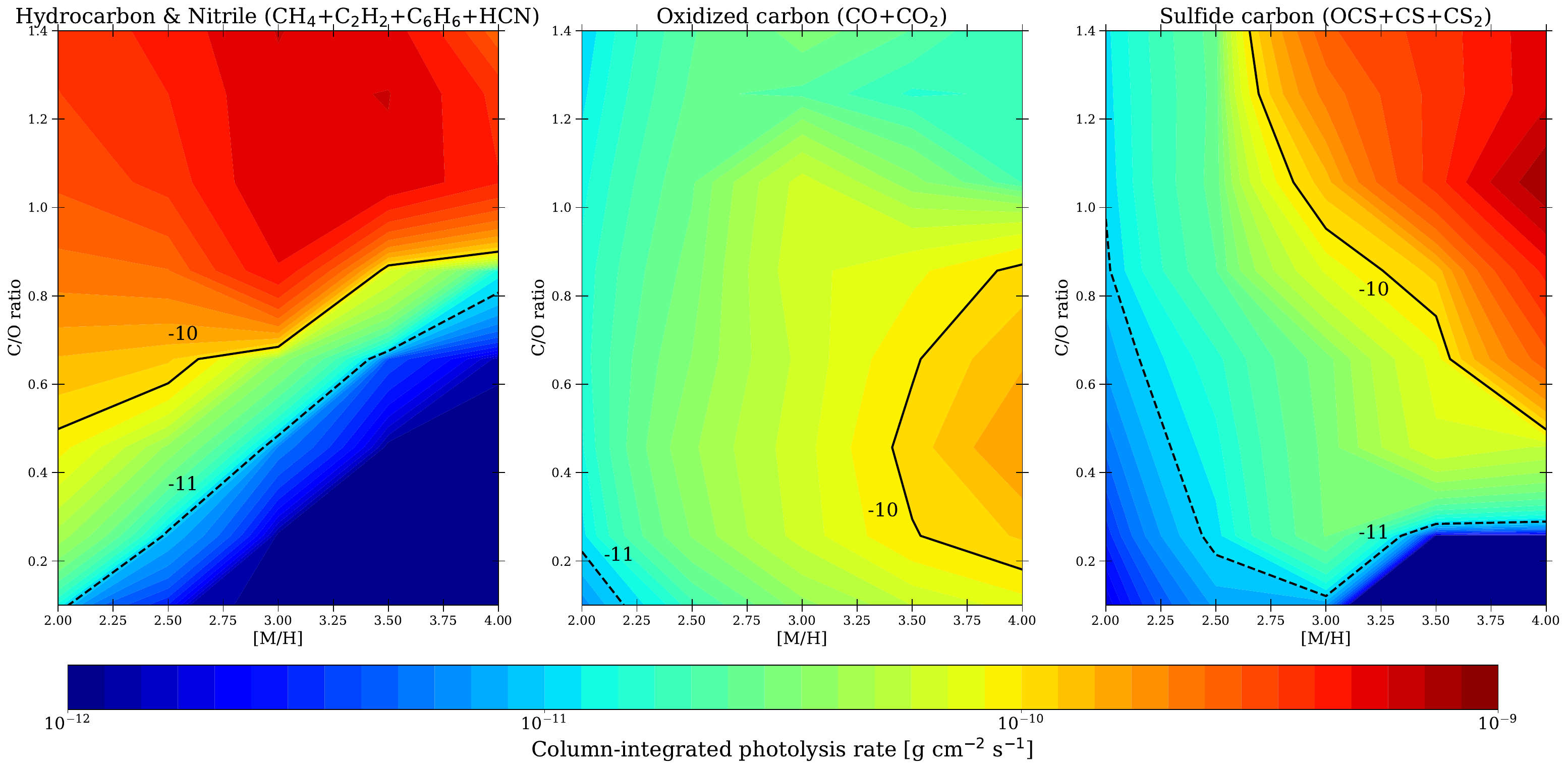}
    \includegraphics[width=\linewidth]{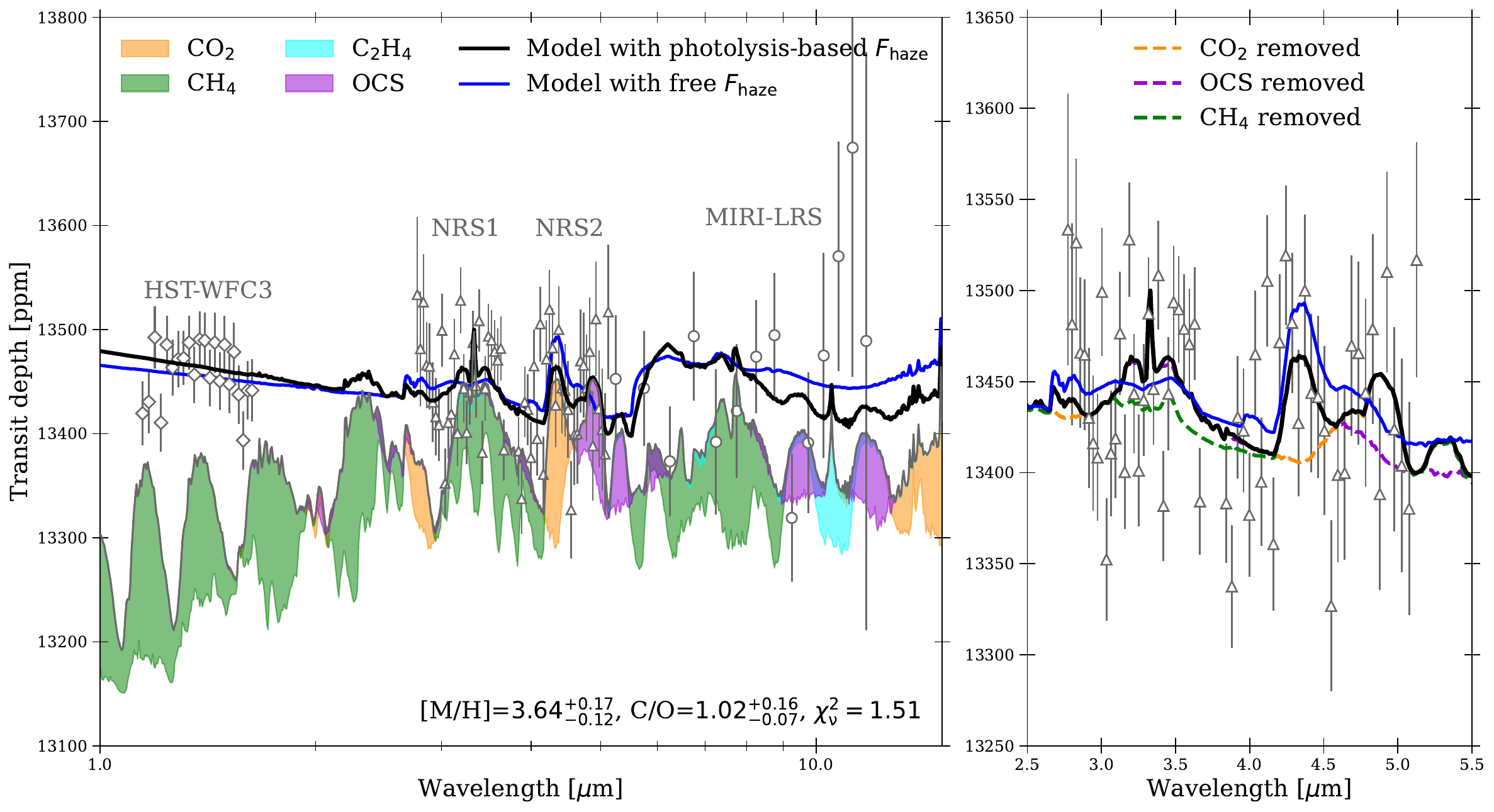}
    \caption{(Top) Column-integrated photolysis rate of hydrocarbons/nitriles (CH$_4$+C$_2$H$_2$+C$_6$H$_6$+HCN, left panel), CO and CO$_2$ (middle panel), and sulfur-based species (OCS+CS+CS$_2$, right panel) as a function of atmospheric metallicity [M/H] and C/O ratio. We assume the 1 bar eddy diffusion coefficient of $K_{\rm zz,1bar}$ in this figure. (Bottom) Same as Figure \ref{fig:median_haze_He400}, but for the median model of PM haze grid retrieval with the photolysis-based $F_{\rm haze}$. The solid blue line shows the median model with the freely-varied $F_{\rm haze}$ for comparison.}
    \label{fig:Fhaze}
\end{figure}

\begin{figure*}
    \centering       
    \includegraphics[width=\linewidth]{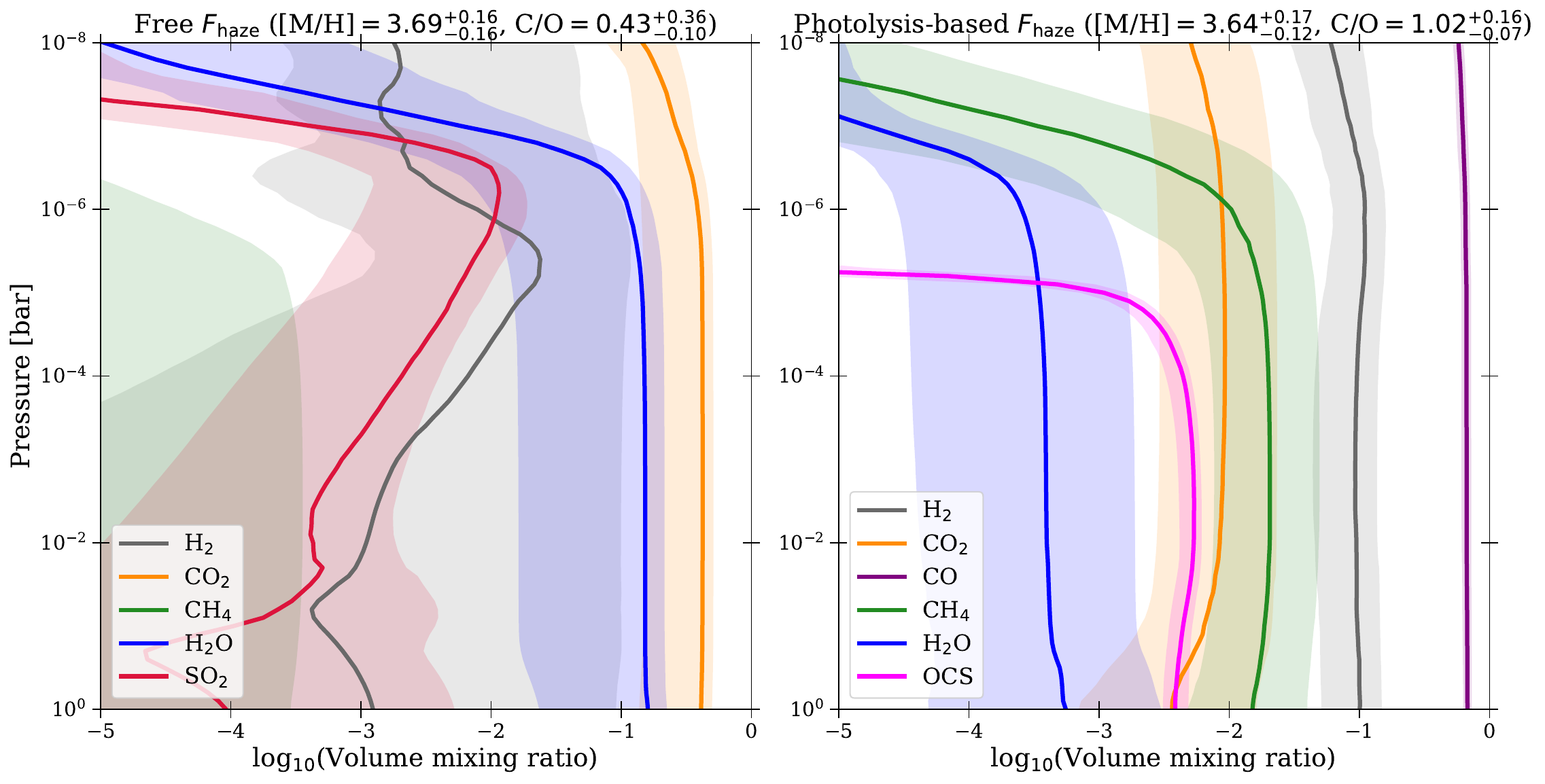}    
    \caption{\rev{Vertical distributions of representative molecules sampled from posterior distributions obtained by photochemical-microphysical (PM) haze grid \rev{with freely-varied $F_{\rm haze}$ (left) and photolysis-based $F_{\rm haze}$ (right).}
    The colored shaded regions denote the 1$\sigma$ confidence interval.}}\label{fig:chem_profile}
\end{figure*}
\rev{Thus far, we have agnostically varied} the haze production rate irrespective of atmospheric properties, since our understanding on haze formation process is incomplete.
As a result, the PM haze grid suggests an extremely high column-integrated haze production rate of $\ga{10}^{-8}~{\rm g~cm^{-2}~s^{-1}}$. 
This result is in agreement with \citet{gao2023gj1214} who also used a haze microphysical model to suggest $F_{\rm haze}>{10}^{-10}~{\rm g~cm^{-2}~s^{-1}}$ from the HST and JWST-MIRI transmission spectrum.
However, although the haze formation process in exoplanets remains unknown, it is unclear whether atmospheric photochemistry is capable of sustaining such a high haze production.
In the Solar System, the haze production rate is estimated to be $3\times{10}^{-14}~{\rm g~{cm}^{-2}~s^{-1}}$ for Titan, $1.2\times{10}^{-14}~{\rm g~{cm}^{-2}~s^{-1}}$ for Pluto, and $\sim2$--$8\times{10}^{-15}~{\rm g~{cm}^{-2}~s^{-1}}$ for Triton \citep[e.g.,][]{McKay+01,Lavvas+11,Gao+17,Ohno+21,Lavvas+21,Gao&Ohno24}.
A much higher haze production rate is expected for GJ1214b, as the planet undergoes much more intense stellar UV photons.
However, previous studies of photochemistry on GJ1214b estimated the haze production rate of $\sim{10}^{-12}~{\rm g~{cm}^{-2}~s^{-1}}$ that is much lower than the rate suggested here.

Here, we calculate the column-integrated photolysis rates of various molecules to examine the potential mass budget available for haze production following previous studies (\citealt{Kawashima&Ikoma19,Lavvas+19photochemicalHazes}, see Appendix \ref{Appendix:photolysis} for more details).
The left top panel of Figure \ref{fig:Fhaze} shows the column-integrated photolysis rates of hydrocarbons and nitriles (CH$_4$+C$_2$H$_2$+C$_6$H$_6$+HCN). 
The column-integrated photolysis rate of each molecule is introduced in Appendix \ref{Appendix:photolysis}.
In general, the photolysis rate of hydrocarbons and nitriles tend to be inefficient at higher metallicity with lower C/O ratios due to the depletion of CH$_4$ and its photochemical products, including C$_2$H$_2$ and HCN.
This implies that hydrocarbon/nitrile-based haze formation, like what happens on Titan, may be inefficient at such a high-metallicity oxidized atmosphere. 
The column-integrate photolysis rate of hydrocarbons and nitriles reaches $\sim{10}^{-11}$--${10}^{-10}~{\rm g~{cm}^{-2}~s^{-1}}$ for the environments of GJ1214b, in agreement with previous studies \citep{Kawashima&Ikoma19,Lavvas+19photochemicalHazes}.
However, the calculated rate appears to be much lower than the rate suggested by the PM haze grid retrieval by approximately two orders of magnitude even at C/O$>1$.

In addition to hydrocarbons and nitriles, we further investigate the column-integrated photolysis rates of CO, CO$_2$, and sulfur-based species (OCS, CS, CS$_2$), as shown in the middle and right top panels of Figure \ref{fig:Fhaze}. 
This exercise is motivated by recent laboratory studies reporting that these molecules serve as haze precursors and/or assist in haze production \citep{He+20,He+20_sulfur}.
We found that the photolysis rate of CO, CO$_2$, and sulfur-based species increases with increasing atmospheric metallicity.
In particular, CO$_2$ and OCS provide high photolysis rates even at high metallicity of [M/H]$\ga3.0$ with C/O$<1$, for which the hydrocarbon/nitrile photolysis tends to be inefficient.
It is also worth noting that OCS photolysis rate is sensitively dependent on eddy diffusion and rapidly increases with increasing $K_{\rm zz}$ (Appendix \ref{Appendix:photolysis}).
These results indicate that the haze formation may have a regime transition from hydrocarbon/nitrile-based process to CO/CO$_2$/sulfur-based process at high metallicity atmospheres.
However, the photolysis rates of these molecules are typically $\sim{10}^{-11}$--${10}^{-10}~{\rm g~{cm}^{-2}~s^{-1}}$, which is still far below the rate of $\ga{10}^{-8}~{\rm g~{cm}^{-2}~s^{-1}}$ suggested by our fiducial model-data comparisons.

Although the actual relation between photolysis and haze production rates is uncertain, the precedent argument raises the concern that our metallicity inference could be biased by an unrealistically high haze production rate.
To assess this, we repeat the PM haze grid retrieval assuming that the column-integrated haze production rate $F_{\rm haze}$ is calculated as the sum of the column-integrated photolysis rate of hydrocarbons (CH$_4$, C$_2$H$_2$, C$_6$H$_6$), nitriles (HCN), CO/CO$_2$, and sulfur-based species (OCC, CS, CS$_2$) for given [M/H], C/O, and $K_{\rm zz,1bar}$.
This model setup allows us to conservatively choose the haze production rate so that it is sustainable by the photolysis of hydrocarbons/nitriles, CO/CO$_2$, and sulfur-containing species.
For this exercise, we adopt the refractive index of water-rich tholin.

The bottom panels of Figure \ref{fig:Fhaze} show the median spectrum of the PM haze grid retrieval with photolysis-based haze production.
The median model could still reasonably fit the current data ($\chi^2_{\rm \nu}=1.51$ for 89 degrees of freedom, 96 datapoints, 7 model parameters) and suggests an extremely high metallicity of ${\rm [M/H]}=3.64^{+0.17}_{-0.12}$, in close agreement with the fiducial PM haze grid that freely varies $F_{\rm haze}$.
This result strengthens our conclusion that GJ1214b likely has a metal-dominated atmosphere.

Although \rev{our metallicity inference is robust against the different prescriptions for the haze production}, we found that the median C/O ratio is considerably different from the model \rev{with freely-varied $F_{\rm haze}$}.
The grid retrieval with photolysis-based $F_{\rm haze}$ obtains a super-solar value of C/O=$1.02^{+0.16}_{-0.07}$, whereas we recall that the fiducial grid retrieval with freely-varied $F_{\rm haze}$ obtains a sub-solar value of C/O=$0.43^{+0.36}_{-0.10}$ (see Figure \ref{fig:differentRetrieval}).
\rev{
The difference in C/O ratio greatly affects the atmospheric molecular abundances for high atmospheric metallicity, as shown in Figure \ref{fig:chem_profile} \citep[see also][]{Hu&Seager14}.
While the freely-varied $F_{\rm haze}$ model infers abundant CO$_2$, H$_2$O, and SO$_2$, the photolysis-based $F_{\rm haze}$ model infers abundant CO, CH$_4$, CO$_2$, and OCS due to the carbon-rich environments.
}
\rev{Thus}, the median spectrum \rev{of the photolysis-based $F_{\rm haze}$ model} produces prominent features of CH$_4$ at $\sim3.2~{\rm {\mu}m}$ and OCS feature at $\sim4.8~{\rm {\mu}m}$ along with CO$_2$ features at $\sim2.7~{\rm {\mu}m}$ and $\sim4.3~{\rm {\mu}m}$.
The photolysis-based $F_{\rm haze}$ model prefers C/O$\sim1$ because it provides high photolysis rates of carbon-bearing species while still guaranteeing a substantial abundance of CO$_2$ whose feature amplitude is comparable to CH$_4$ feature.
Since the haze production rate is restricted to $F_{\rm haze}\sim{10}^{-9.5}~{\rm g~cm^{-2}~s^{-1}}$ in this model setup, the haze absorption feature at $\sim3$--$3.6~{\rm {\mu}m}$ is no longer visible. 
Instead, the CH$_4$ feature shapes the spectrum in this wavelength range.
It remains unclear which sub-solar or super-solar C/O solution is more likely from the NIRSpec spectrum given the current signal-to-noise ratio and the uncertain offset between the NRS1 and NRS2 detectors (see Appendix \ref{Appendix:offset} for futrher discussions).


\subsection{Summary of Model--Data Comparison: A Metal-Dominated Atmosphere on GJ1214b?}

\begin{figure}
    \centering       
    \includegraphics[width=0.80\linewidth]{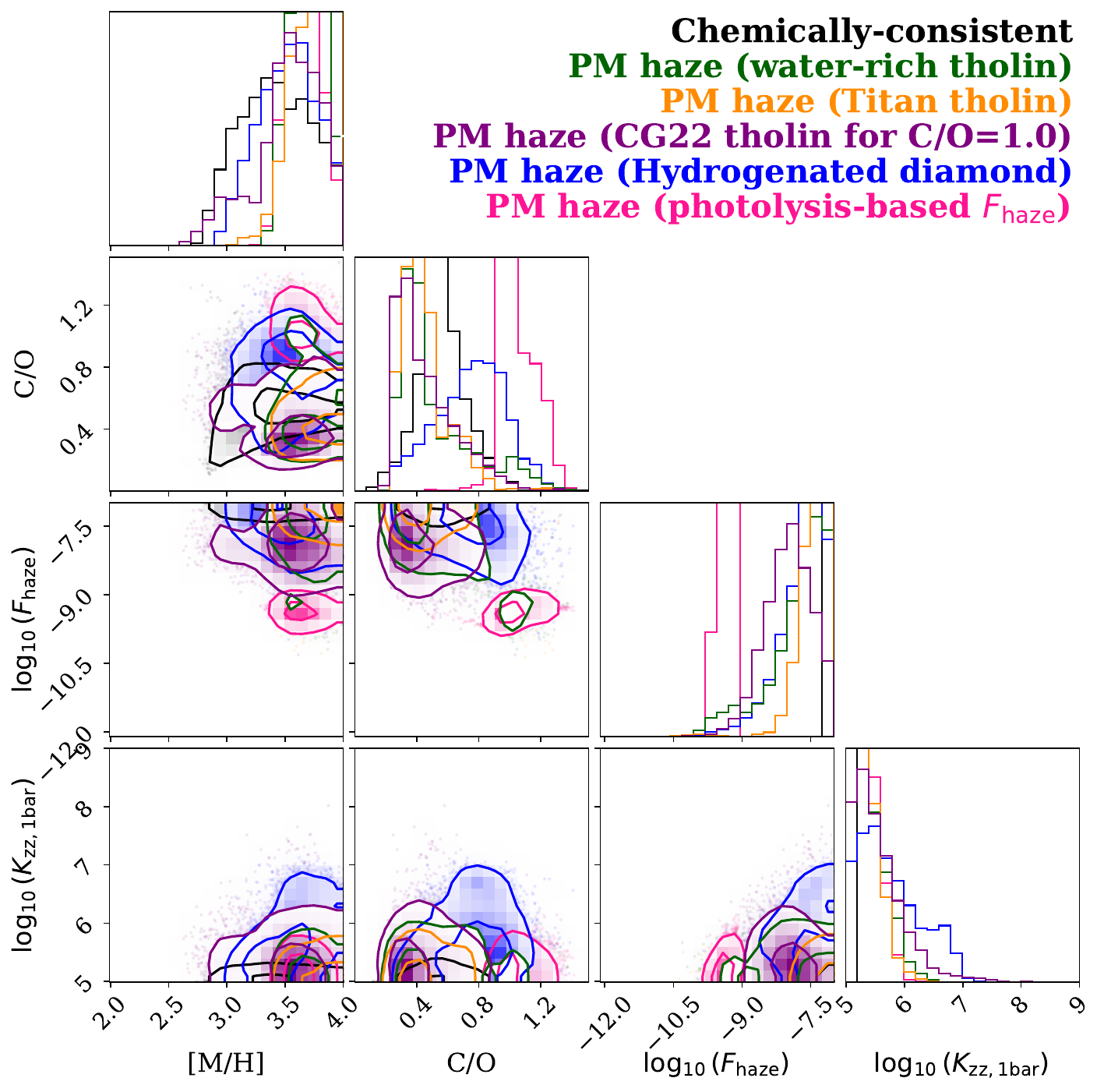}   
\centering
\begin{tabular}{l||c|c|c|c|c} 
    \hline \hline
    Model configuration & [M/H] & C/O & $\log_{\rm 10}(F_{\rm haze})$ & $\log_{\rm 10}(K_{\rm zz,1bar})$ & $\chi^2/N_{\rm data}$ \\ \hline \hline
    PM haze grid (Water-rich tholin) & $3.69^{+0.16}_{-0.16}$   & $0.43^{+0.36}_{-0.10}$ & $-7.61^{+0.41}_{-0.77}$ & $5.34^{+0.35}_{-0.23}$ & 1.36 \\ \hline
    PM haze grid (Titan tholin) & $3.77^{+0.16}_{-0.22}$   & $0.42^{+0.19}_{-0.10}$ &  $-7.31^{+0.21}_{-0.32}$ & $5.26^{+0.28}_{-0.18}$ & 1.42 \\ \hline
    PM haze grid (CG22 tholin for C/O$=1.0$) & $3.57^{+0.24}_{-0.31}$   & $0.41^{+0.26}_{-0.12}$ &  $-7.95^{+0.50}_{-0.54}$ & $5.45^{+0.48}_{-0.29}$ & 1.35 \\ \hline
    PM haze grid (Hydrogenated diamond) & $3.54^{+0.26}_{-0.28}$   & $0.77^{+0.20}_{-0.25}$ &  $-7.58^{+0.35}_{-0.56}$ & $5.65^{+0.71}_{-0.42}$ & 1.28 \\ \hline
    PM haze (Water-rich tholin/photolysis-based $F_{\rm haze}$) & $3.64^{+0.17}_{-0.12}$   & $1.02^{+0.16}_{-0.07}$ &  \rev{($-9.34^{+0.09}_{-0.17}$)} & $5.30^{+0.27}_{-0.20}$ & 1.42 \\ \hline
    Chemically-consistent retrieval & $3.48^{+0.42}_{-0.36}$ & $0.55^{+0.15}_{-0.12}$ & N/A & N/A & 1.53 \\ 
  \end{tabular}

\caption{ 
    (Top) Corner plots showing the 2D posterior distributions of atmospheric metallicity [M/H], C/O ratio, column-integrated haze production rate $F_{\rm haze}~{\rm [g~cm^{-2}~s^{-1}]}$, and $1~{\rm bar}$ eddy diffusion coefficient at $K_{\rm zz,1bar}~{\rm [cm^{2}~s^{-1}]}$ for PM haze grid retrieval with various choice of haze refractive indices. The figure also overplots the posteriors of [M/H] and C/O obtained by the chemically-consistent retrieval. (Bottom) Summary of median values with 1$\sigma$ confidence interval of key model parameters for each retrieval configuration. \rev{For the photolysis-based $F_{\rm haze}$ model, the table denotes the value of $\log_{\rm 10}(F_{\rm haze})$ calculated from the posterior samples of [M/H], C/O, and $\log_{\rm 10}(K_{\rm zz})$.}
    } 
    \label{fig:differentRetrieval}
\end{figure}

To further assess the sensitivity of our interpretation to the aerosol assumptions, we utilize the result of chemically-consistent retrieval that was adopted in \citet{schlawin2024gj1214b_placeholder} to obtain the initial fit to the NIRSpec/G395H transmission spectrum.
This retrieval used the \texttt{CHIMERA} code to calculate transmission spectrum, where the TP profile is parameterized by the analytic model of \citet{guillot2010radEquilibrium}, and the aerosol opacity is parameterized by a power-law function $\kappa_{\rm aer}=\kappa_{\rm 0}(\lambda/\lambda_{\rm 0})^{-\alpha}$ without assuming a specific aerosol type.
The molecular abundances are calculated by \texttt{FastChem2} \citep{stock2022fastchem2} for thermochemical equilibrium, and the model also introduces the disequilibrium effects following the prescription of \citet{morley2017gj436bRetrieval}.
We refer the reader to \citet{schlawin2024gj1214b_placeholder} for further details.
Figure \ref{fig:differentRetrieval} shows the 2D posterior distributions of atmospheric metallicity and C/O obtained by the chemically-consistent retrieval.
We also overplot the posteriors from the PM haze grid retrieval for various haze optical constants tested in Section \ref{sec:many_optic} \rev{}.
The chemically-consistent retrieval obtains the atmospheric metallicity of ${\rm [M/H]}=3.48^{+0.42}_{-0.36}$ and C/O$=0.55^{+0.15}_{-0.12}$. 
These results agree well with the results of the PM haze grid retrieval despite considerably different model setups.

The atmospheric metallicity of [M/H]$\sim3.5$--$3.8$ is equivalent to the atmosphere where hydrogen is no longer the main constituent.
\rev{Figure \ref{fig:chem_profile}} demonstrates that volume mixing ratios of \rev{either CO$_2$, CO, or H$_2$O} exceed H$_2$, i.e., partial pressure of heavy molecules is higher than that of hydrogen.
Interestingly, \rev{for the sub-solar C/O ratio,} H$_2$ is no longer the main hydrogen reservoir in such a metal-dominated atmosphere and instead a photochemical product, as similar to H$_2$ in Titan's atmosphere \citep{Strobel2010_titan_H2}.

\section{Summary and Discussion}\label{sec:summary}

We have investigated the atmospheric properties of GJ1214b based on its panchromatic transmission spectrum established by HST/WFC3,  JWST/NIRSpec and JWST/MIRI.
We have used a suite of atmospheric models, including photochemical and aerosol microphysical models to constrain the atmospheric compositions. 
The photochemical-microphysical (PM) haze model has suggested an extremely high metallicity of [M/H]$\sim3.5$--$3.8$ regardless of the choice of the haze optical constants.
Our analysis has shown that the spectrum of each HST/WFC3, JWST/G395H-NRS1, JWST/G395H-NRS2, and JWST/MIRI-LRS instrument individually prefers such high metallicity. 
The extreme metallicity is supported by the relatively strong CO$_2$ feature compared to haze and/or CH$_4$ features in the NIRSpec spectrum, as well as the flattness of the MIRI spectrum.
We have tested a number of assumptions for uncertain haze properties, including refractive indices and production rate, and chemically-consistent retrieval that is agnostic for the type of aerosol.
All the models tested here have achieved a consistently high metallicity of [M/H]$\sim3.5$.
Consequently, we conclude that the current transmission spectrum indicates a metal-dominated atmosphere on GJ1214b where hydrogen is no longer the main atmospheric constituent.
\rev{A contemporaneous work by \citet{Lavvas+24} also reached the similar conclusion of $2000$--$3000\times$ solar metallicity for the GJ1214 b atmosphere based on its HST and MIRI transmission and emission spectra.}
This study has independently supported the high metallicity inferred by the MIRI phase curve \citep{kempton2023reflectiveMetalRichGJ1214} and by previous modeling studies of aerosol formation \citep{morley2015superEclouds,Charnay+15GJ1214bspectrum,Ohno&Okuzumi18microphysicalCloudModelsGJ214gj436b,gao2018microphysicsGJ1214clouds,Kawashima+19,Lavvas+19photochemicalHazes,Ohno+20fluffyaggregate,gao2023gj1214}.
GJ1214b joins a growing group of super-Earths/sub-Neptunes with extremely metal-rich atmospheres, including L98-59d \citep{Banerjee+24_L98-59d_H2S,Gressier+24}, GJ9827d \citep{Piaulet+24}, and LTT9779b \citep{hoyer2023LTT9779b}.

The possible metal-dominated atmosphere of GJ1214b provides many implications for the formation and evolution of sub-Neptunes.
During planet formation, solid (e.g., planetesimals, pebbles) accretion followed by ablation/sublimation causes metal enrichment in protoatmospheres \citep[e.g.,][]{Pollack+86,fortney2013lowMassP,Venturini16_enrichment,Shibata&Helled22}.
Giant impacts with high impact energy could also dredge up the vaporized core into the atmosphere to cause metal enrichment \citep{Kurosaki&Inutsuka23}.
Accretion of vapor-enriched disk gas also forms metal-rich atmospheres \citep[e.g.,][]{Booth+17,Schneider&Bitsch21}, but this process alone achieves atmospheric metallicity only up to $\sim10$--$30\times$ solar value \citep{Danti+23}, far below the metallicity suggested for GJ1214b.
The post-formation process can also induce metal enrichment in the atmosphere. 
For instance, magma-atmosphere interaction \citep[e.g., H$_2$O production from reaction of H$_2$ with oxidized magma,][]{Ikoma&Genda06} can alter atmospheric compositions; however, this process can achieve the mean molecular weight up to $<7~{\rm amu}$ \citep{Kite+20} and results in very low atmospheric C/O ratio of $\ll0.1$ \citep{Seo+24}, not compatible with the metallicity and C/O ratio inferred for GJ1214b.
The degassing of the rocky mantle could produce CO$_2$ dominated atmospheres for proper combinations of the mantle oxidization state and the carbon contents \citep{Tian&Heng24}, where the carbon content is related to how much refractory carbons, such as CHON organics, are contained in the planetary building blocks \citep{Bergin+23}.
However, a rocky core surrounded by a CO$_2$ dominated atmosphere is in tension with the mass-radius relation of GJ1214b, which requires a substantial amount of hydrogen in the atmosphere to make a rocky core being consistent with planet's low density \citep[e.g.,][]{valencia2013bulkCompositionGJ1214b,Nixon+24internalStructureGJ1214bJWST}.
Based on these arguments, the metal-dominated atmosphere of GJ1214b possibly originates from planetesimal/pebble accretion during the formation; otherwise, giant impact events might be responsible for the metal-dominated atmosphere.


Although we mentioned that a rocky core is in tension with mass-metallicity relation above, a metal-dominated atmosphere still poses a challenge for the current interior structure model of sub-Neptunes even if the planet's interior contains ice. 
The latest interior structure model showed that GJ1214b needs a nearly pure H$_2$O core to explain the mass-radius relation if the planet has a H$_2$O steam atmosphere \citep{Nixon+24internalStructureGJ1214bJWST}.
This study suggests that GJ1214b has a CO$_2$-dominated atmosphere with a higher mean molecular weight, making the envelope even thinner and thus it is harder to explain the planet's radius.
The seemingly too metal-rich atmosphere of GJ1214b may suggest that previously ignored processes play a vital role in controlling the present day radii of sub-Neptunes, such as enhanced atmospheric opacity that slows down thermal evolution \citep{Burrows+07opacity}, energy release by metal rainout \citep{Vazan+24silicateRainoutDelaysContractionSubNeptunes} and radiative feedback from aerosols \citep{Poser&Redmer24cloudythermalEvolution}.
Although conventional interior structure models used only H$_2$O as ''ice``, astronomical ices contain a non-negligible amount of CO and CO$_2$ ices \citep{Boogert+15_ISM_composition,McClure+23_iceage} that might affect the interior structure of icy sub-Neptunes if they originally formed beyond CO/CO$_2$ snowlines.

This study has postulated that photochemical haze is the predominant aerosol in the GJ1214b atmosphere but does not rule out the possibility of other type of aerosols.
Condensed KCl clouds are the leading candidate for the thick aerosol layer on GJ1214b \citep{Morley2013_gj1214b,Charnay+15GJ1214bspectrum,Ohno&Okuzumi18microphysicalCloudModelsGJ214gj436b,gao2018microphysicsGJ1214clouds,Ohno+20fluffyaggregate,Christie2022_gj1214b_gcm}.
\citet{Huang+24} also recently showed that Na$_2$S clouds ascending from a deep atmosphere can also be responsible for flattening the GJ1214b spectrum if the eddy diffusion coefficient is extremely high.
Since condensed minerals, especially KCl, tend to have reflective optical properties, reminiscent of the putative ``reflective aerosol'' suggested by \citet{kempton2023reflectiveMetalRichGJ1214}, it would be worthwhile to investigate this possibility.
Although we have assumed spherical haze particles for simplicity, aerosol particles with fractal aggregates could affect the spectral shape and potentially the inferred atmospheric properties \citep{Adams+19aggregateHazes,Ohno+20fluffyaggregate}.
Since the difference in aerosol nature likely affects the atmospheric thermal structure, future modeling studies would be worthwhile to investigate what aerosol scenario provides a unified way to explain all existing observations, including the transmission spectrum \citep{kreidberg14,kempton2023reflectiveMetalRichGJ1214,schlawin2024gj1214b_placeholder}, dayside/nightside emission spectra and spectroscopic phase curve \citep{kempton2023reflectiveMetalRichGJ1214}.

We also stress the importance of follow-up observations of GJ1214b.
The signal-to-noise ratio of the observed spectral features is still not very high; for example, \citet{schlawin2024gj1214b_placeholder} reported that the detection significance of CO$_2$ and CH$_4$ are only $2.4\sigma$ and $2.0\sigma$, respectively. 
The current NIRSpec spectrum also struggles to distinguish the model with different C/O ratios with different assumptions for the haze production (Section \ref{discussion:Fhaze}).
Follow-up observations by NIRSpec are highly warranted to make a possible metal-dominated atmosphere compelling and to better constrain the atmospheric C/O ratio, both of which provide crucial implications for the formation, evolution and interior structures of sub-Neptunes.
Follow-up observations by JWST-MIRI would also be worthwhile. 
The flatness of the MIRI spectrum suggests a small scale height with a high metallicity, whereas the haze particles are expected to produce an organic feature at $\sim6~{\rm {\mu}m}$ with $\sim50~{\rm ppm}$ amplitude.
Follow-up high precision MIRI observations would tightly constrain the atmospheric scale height and potentially provide the opportunity to directly observe the organic feature of exoplanetary hazes for the first time, which would allow us to diagnose what kind of aerosols are actually veiling the atmosphere of GJ1214b.


\begin{acknowledgements}
K.Ohno acknowledge funding support from the JSPS KAKENHI Grant Number JP23K19072 and JP21H01141.
Funding for E. Schlawin is provided by NASA Goddard Spaceflight Center.
We respectfully acknowledge the University of Arizona is on the land and territories of Indigenous peoples. 
T.J.B. and T.P.G.~acknowledge funding support from the NASA Next Generation Space Telescope Flight Investigations program (now JWST) via WBS 411672.07.04.01.02 and 411672.07.05.05.03.02.
This work benefited from the 2024 Exoplanet Summer Program in the Other Worlds Laboratory (OWL) at the University of California, Santa Cruz, a program funded by the Heising-Simons Foundation and NASA.
A part of numerical computations were carried out on PC cluster at Center for Computational Astrophysics, National Astronomical Observatory of Japan.
\end{acknowledgements}

\rev{The JWST data presented in this Letter were obtained from the Mikulski Archive for Space Telescopes (MAST) at the Space Telescope Science Institute. 
The specific observations analyzed can be accessed via DOI 10.17909/tqa7-th94 and are accessible in \citep[][DOI: 10.3847/2041-8213/ad7fef]{schlawin2024gj1214b_placeholder}.
}

%

\vspace{5mm}


\software{astropy \citep{astropy2013}, 
          \texttt{pysynphot} \citep{lim2015pysynphot},
          \texttt{matplotlib} \citep{Hunter2007matplotlib},
          \texttt{numpy} \citep{vanderWalt2011numpy},
          \texttt{scipy} \citep{virtanen2020scipy},
          \texttt{CHIMERA} \citep{line2013chimera},          
          \texttt{PyMieScatt} \citep{Sumlin+18},
          \texttt{PyMultiNest} \citep{Buchner+14_pymultinest},
          \texttt{VULCAN}
          \citep{tsai2021VULCANcomparitiveStudy},
          \texttt{FastChem2} \citep{stock2022fastchem2}
           }



\appendix

\section{Photochemical-Microphysical Haze Model}\label{sec:appendix_method1}

\subsection{Atmospheric Thermal, Chemical, and Haze Profiles}
\begin{figure}
    \centering
    \includegraphics[width=\linewidth]{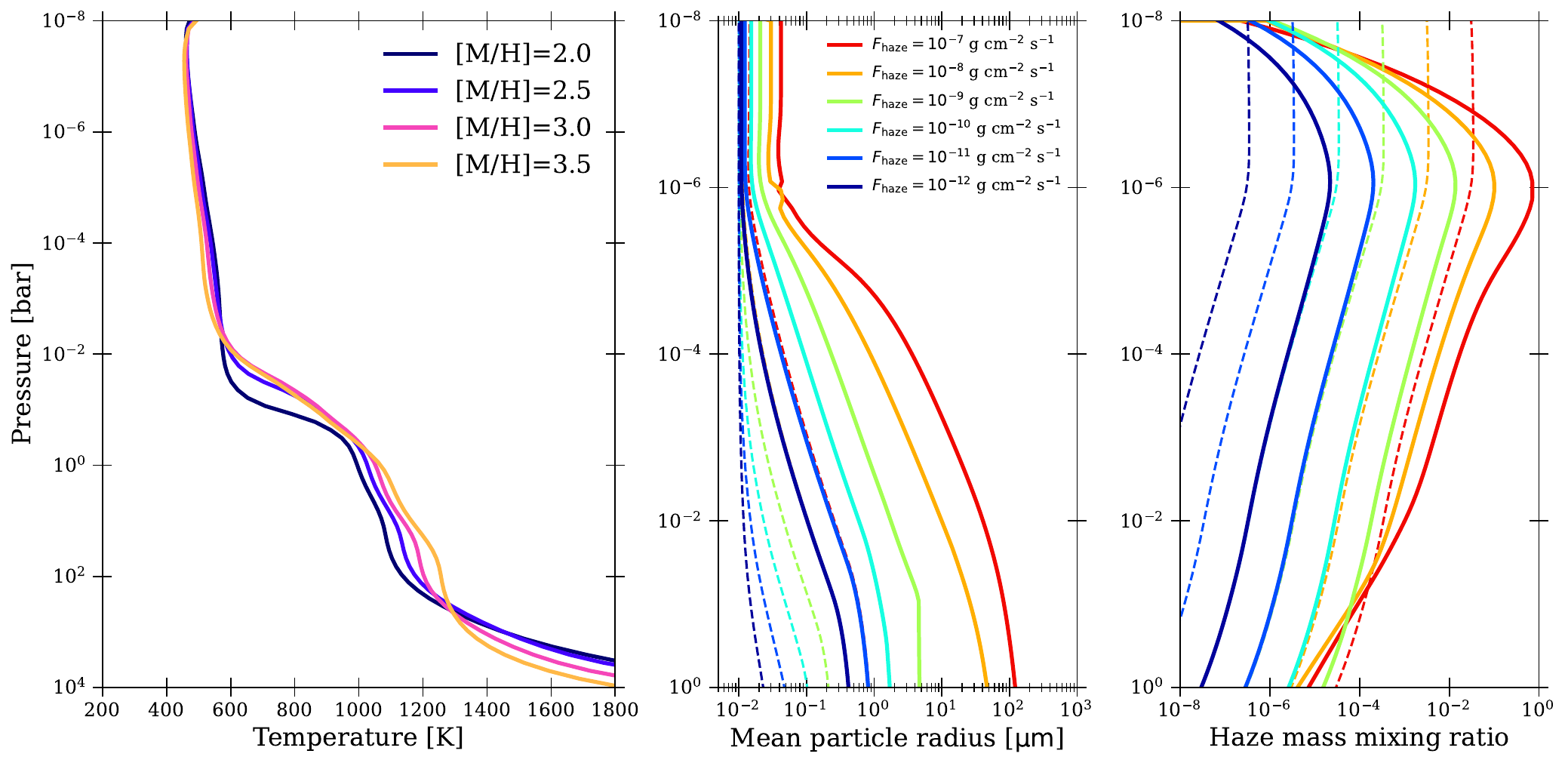}
    \caption{Example of TP and haze profiles. In the left panel, different colored lines show the vertical TP profiles for different atmospheric metallicities used in this study. The middle and right panels show the vertical distributions of mean radius and mass mixing ratio of haze particles. Different colored lines show the profiles for different column-integrated haze production rate $F_{\rm haze}$. The solid and dashed lines show the profiles for $K_{\rm zz,1bar}=10^6$ and $10^8~{\rm cm^2~s^{-1}}$, respectively.
    }
    \label{fig:TP_haze}
\end{figure}

We first compute the atmospheric temperature-pressure (TP) profile with a radiative-convective equilibrium model \texttt{EGP} used by many previous studies \citep[e.g.,][]{Marley&McKay99thermalStructure,Fortney+05comparitivePlanetaryAtmospheres,Morley2013_gj1214b,marley2015rev,Thorngren2019intrinsicTemperature,gao2020aerosolsSilicatesAndHazes,OhnoFortney2021_Nitrogen1}.
The input stellar spectrum of GJ1214 is extracted from the phoenix stellar grid compiled in the \texttt{PySynphot} package \citep{lim2015pysynphot} for $T_{\rm eff}=3250~{\rm K}$, [Fe/H]=0.29, and $\log{(g)}=5.026$ \citep{cloutier2021gj1214bPreciseMass}.
We calculated the TP profiles for the atmospheric metallicity of [M/H]=2.0, 2.5, 3.0, and 3.5 for the surface gravity of $g=10.65~{\rm m~s^{-2}}$ \citep{cloutier2021gj1214bPreciseMass} assuming full heat redistribution.
Although the MIRI phase curve shows strong day-night temperature contrast in GJ1214b \citep{kempton2023reflectiveMetalRichGJ1214}, the full heat redistribution would still be a reasonable assumption for transmission spectroscopy that probes intermediate terminator regions.
We assume the planetary intrinsic temperature of $T_{\rm int}=40~{\rm K}$ that is consistent with the thermal evolution models for GJ1214b \citep{Nettelmann2011gj1214,valencia2013bulkCompositionGJ1214b}.
The left panel of Figure \ref{fig:TP_haze} shows the calculated TP profiles.

The computed TP profiles are post-processed in the microphysical and photochemical models.
To investigate the effects of haze on observable atmospheric spectra, we created a grid of photochemical haze profiles using a microphysical model of \citet{Ohno&Okuzumi18microphysicalCloudModelsGJ214gj436b} and \citet{Ohno+20fluffyaggregate} that was extended to photochemical hazes in \citet{Ohno&Kawashima20superRaleighSlopes}. 
The model adopts a two-moment scheme that calculates the vertical distributions of the number and mass densities of haze particles, $n_{\rm p}$ and $\rho_{\rm p}$, at each pressure level by taking into account collision growth, vertical transport by eddy diffusion, and gravitational settling.
The number and mass densities are then converted to the mean particle radius from the relation of $\rho_{\rm p}/n_{\rm p}=4\pi \rho_{\rm 0}r_{\rm p}^3/3$, where $\rho_{\rm 0}$ is the internal density of the haze particles and set to $\rho_{\rm p}=1~{\rm g~{cm}^{-3}}$.
The moment scheme enables a fast computation to achieve the vast parameter survey.
Our approach closely follows \citet{Ohno&Kawashima20superRaleighSlopes}.
The model injects initial seed haze particles with a vertical profile of particle production rate that obeys the lognormal distribution centered at $P={10}^{-6}~{\rm bar}$ with a width of $\sigma=0.5$. 
The model then calculates the subsequent collision growth and the vertical transport of the haze particles through eddy diffusion and gravitational settling, where we have assumed compact spherical particles for simplicity.
The eddy diffusion coefficient is assumed to obey the following functional form:
\begin{equation}
    K_{\rm zz}=K_{\rm zz,1bar}\left( \frac{P}{1~{\rm bar}}\right)^{-0.4},
\end{equation}
where the pressure dependence is motivated by a previous GCM study of \citet{Charnay15_gcm}.
For reference, \citet{Charnay15_gcm} obtained $K_{\rm zz,1bar}={7\times{10}^{6}}$, $2.8\times{10}^{7}$, $3\times{10}^{7}$, and $3\times{10}^{6}~{\rm cm^2~s^{-1}}$ for the 1, 10, 100$\times$ solar metallicity and pure water atmospheres for GJ1214b.
In this study, $K_{\rm 1bar}$ is treated as a free parameter.
For TP profiles of each atmospheric metallicity, we calculate the vertical distributions of the mean particle size and the mass mixing ratio of photochemical hazes for finely spaced grids of the column-integrated haze production rate of $F_{\rm haze}={10}^{-7}$--${10}^{-12}~{\rm g~cm^{-2}~s^{-1}}$ and 1 bar eddy diffusion coefficient of $K_{\rm 1bar}={10}^{5}$--${10}^{9}~{\rm cm^2~s^{-1}}$ at every 0.2 dex.
In total, we calculated 2184 haze profiles to constrain the haze properties from the panchromatic spectrum.

The middle and right panels of Figure \ref{fig:TP_haze} show the example of the vertical distributions of the mean radius and the mass mixing ratio of haze particles for various $F_{\rm haze}$ at $K_{\rm zz,1bar}={10}^{6}$ and ${10}^{8}~{\rm cm^2~s^{-1}}$. 
In general, particle radius becomes larger at lower atmosphere because of collision growth, and a higher $F_{\rm haze}$ leads to larger particles with a higher mass mixing ratio.
Meanwhile, stronger eddy diffusion reduce the particle radius and mass mixing ratio by efficiently removing haze particles from the upper atmosphere.
We refer the reader to previous studies on haze modeling for a more in-depth analysis of the haze structure \citep[e.g.,][]{Kawashima&Ikoma19,Lavvas+19photochemicalHazes,Ohno&Kawashima20superRaleighSlopes,gao2023gj1214}.

For chemical profiles, we use the open-source photochemical model \texttt{VULCAN} \citep{tsai2021VULCANcomparitiveStudy} to calculate the vertical distributions of the molecular abundances.
We adopted the SNCHO chemical network that includes 571 forward reactions and their reversed reactions (1142 reactions in total) with supplemented by photodissociation reactions of 69 species.
The input stellar UV spectrum for GJ1214 is taken from the MUSCLE survey\footnote{\url{https://archive.stsci.edu/prepds/muscles/}} \citep{France+16_MUSCLE,Younblood+16_MUSCLE,Loyd+16_MUSCLE}.
For each atmospheric metallicity of [M/H]=2.0, 2.5, 3.0 and 3.5 with the solar abundance of \citet{lodders2009abundances}, we run the photochemical model for possible combinations of eddy diffusion coefficient ($K_{\rm zz,1bar}={10}^{5}$, ${10}^{6}$, ${10}^7$, ${10}^8$, ${10}^{9}~{\rm cm^2~s^{-1}}$) and the carbon-to-oxygen ratio (C/O=$0.057$, $0.257$, $0.457$, $0.657$, $0.857$, $1.057$, $1.257$, $1.457$), where N/O and S/O ratios are fixed to the solar value.
During the preparation of this manuscript, we found that the metallicity inferred converges toward the upper limit of [M/H]$\sim3.5$.
We therefore extend the grid to [M/H]$=4.0$ by assuming that the TP profile of [M/H]$=4.0$ is approximately the same as that of [M/H]$=3.5$.
In total, we calculated 200 vertical profiles of molecular abundances based on photochemical calculations to interpret the panchromatic spectrum.

\begin{figure}
    \centering
    \includegraphics[width=\linewidth]{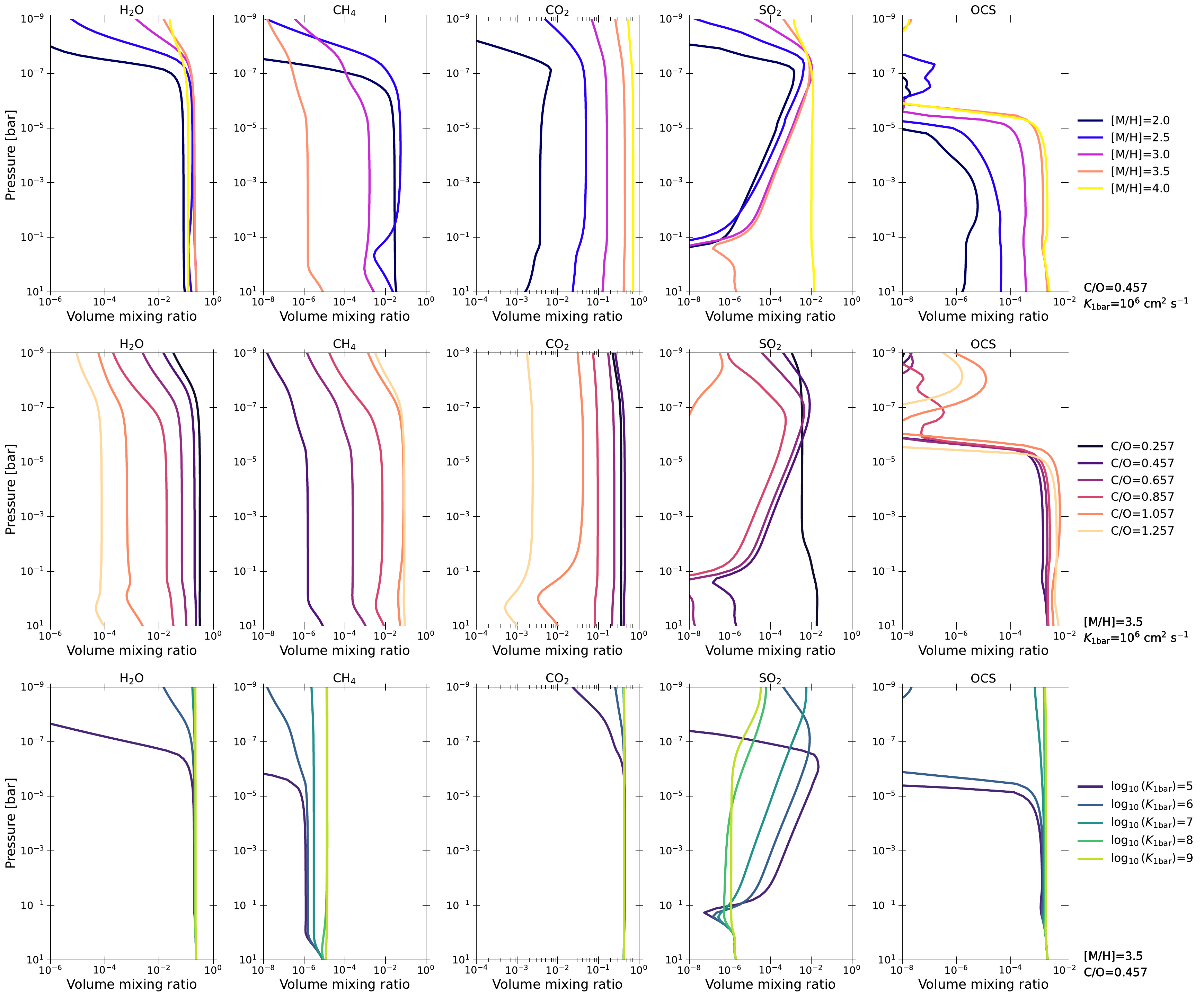}
    \caption{Vertical distributions of several key molecules for various metallicity [M/H], C/O ratio, and eddy diffusion coefficient. From left to right, each column shows the volume mixing ratio of H$_2$O, CH$_4$, CO$_2$, SO$_2$ and OCS at each pressure level, respectively. From top to bottom, different colored lines in each row show the effects of varying [M/H], C/O ratio, and $K_{\rm zz}$, respectively. We use [M/H]=3.5, C/O=0.457, $K_{\rm 1bar}={10}^{6}~{\rm cm^2~s^{-1}}$ as a fiducial parameter set, and then vary one of them to illustrate the parameter dependence. In the bottom row, $K_{\rm 1bar}$ is in cgs unit.}
    \label{fig:VULCAN}
\end{figure}

To clarify the parameter dependences of molecular abundances, Figure \ref{fig:VULCAN} shows the vertical profiles of H$_2$O, CH$_4$, CO$_2$, SO$_2$, and OCS.
H$_2$O abundance is mostly controlled by the C/O ratio and decreases with increasing C/O.
CH$_4$ abundance is sensitive to both [M/H] and C/O. 
For GJ1214b conditions, CH$_4$ abundance typically decreases with increasing [M/H] while increasing with increasing C/O.
CO$_2$ shows the opposite trend to CH$_4$.
CO$_2$ abundance rapidly increases with increasing [M/H] while gradually decreases with increasing C/O.
These results are in line with previous studies \citep[e.g.,][]{moses13}.
SO$_2$ abundance also increases with increasing metallicity, while higher C/O ratio tends to suppress SO$_2$ formation \citep{Tsai2023_SO2,Polman23,Crossfield23,beatty24gj3470b}.
It is interesting to note that SO$_2$ is no longer a photochemical product, and instead the thermochemical equilibrium with vertical quench defines the SO$_2$ abundance for extremely metallicity of [M/H]$\ge3.5$ with low C/O ratio.
The abundance of OCS is known to be sensitive to metallicity in thermochemical equilibrium \citep{moses13} and is also proposed as an indicator of a high H$_2$O abundance in the deep interiors of temperate sub-Neptunes \citep{Yang&Hu24}.
We also see a rapid increase in OCS abundance as the metallicity increases.
A higher C/O ratio also tends to enhance the OCS abundance, and the OCS abundance takes a maximum at C/O$\sim1.0$.
As compared to [M/H] and C/O, the eddy diffusion coefficient has a relatively minor role in controlling the molecular abundance for the GJ1214b environment.
Yet, SO$_2$ and OCS abundances are still sensitive to eddy diffusion.
A stronger eddy diffusion reduces the SO$_2$ abundance while it allows OCS to survive against photodissociation and destruction by atomic hydrogen attack, leading to drastically increase OCS abundance at $P\la{10}^{-5}~{\rm bar}$.

\subsection{Transmission Spectrum and Procedure of Model-Data Comparisons}
We use the open-source radiative transfer model \texttt{CHIMERA} \citep{line2013chimera,Mai&Line19cloudAssumptions} to compute the transmission spectrum of GJ1214b.
The present model includes molecular absorption opacity of H$_2$O \citep{Polyansky+18}, CO \citep{freedman2014opacities}, CO$_2$ \citep{freedman2014opacities},  NH$_3$ \citep{freedman2014opacities}, HCN \citep{freedman2014opacities}, H$_2$S \citep{freedman2014opacities}, CH$_4$ \citep{Rothman2010_hitemp}, C$_2$H$_2$ \citep{Chub2020_exomol_c2h2}, C$_2$H$_4$ \citep{Mant2018_c2h4_exomol}, CH$_3$ \citep{Adam+19}, C$_2$H$_6$ \citep{lupu2022_c2h6_xsection}, CS \citep{Paulose15_exomol_cs}, SO \citep{Brady2024_exomol_so}, SO$_2$ \citep{Underwood2016_exomol_so2}, and OCS \citep{Owens2024_exomol_ocs} using the correlated-k method.
The model also includes the H$_2$-H$_2$ and H$_2$-He collision-induced absorption \citep{Richard2012} and Rayleigh scattering by H$_2$ and He.
For computing the opacity of haze particles, we use the Mie theory code \texttt{PyMieScatt} \citep{Sumlin+18} assuming spherical haze particles.
We adopt the complex refractive indices of tholin synthesized from $1000\times$ solar metallicity gas at $400~{\rm K}$ \citep[water-rich tholin,][]{He+23organichazesWaterRich} as a fiducial case and test other refractive indices of possible haze analogs for exoplanets in Section \ref{sec:many_optic}.
We assume a stellar radius of $0.215~{\rm R_{\rm sun}}$ and a planetary mass of $8.17~{\rm M_{\rm Earth}}$ \citep{cloutier2021gj1214bPreciseMass} to compute transit depth to be compared with the observed spectrum.
We note that these are consistent with the latest values derived from JWST transit lightcurve fitting for the system of $0.2162^{+0.0025}_{-0.0024} ~{\rm R_{\rm sun}}$ for the host star and $8.41^{+0.36}_{-0.35}~{\rm M_{\rm Earth}}$ for planet b \citep{Mahajan+24_gj1214update}.

We compare the theoretical transmission spectrum with the transmission spectrum of GJ1214b observed by HST-WFC3 \citep{kreidberg14}, JWST/NIRSpec-G395H \citep{schlawin2024gj1214b_placeholder}, and JWST/MIRI-LRS \citep{kempton2023reflectiveMetalRichGJ1214} that was renalyzed by \citet{schlawin2024gj1214b_placeholder} using the \texttt{PyMultinest} \citep{feroz2009multinest,Buchner+14_pymultinest} to obtain the posterior distributions for each input parameter.
Through model fitting, we vary the planetary reference radius at $10~{\rm bar}$ $R_{\rm 10bar}$ that defines gravity and thus scale height at each pressure level, atmospheric metallicity [M/H], C/O ratio, column-integrated haze production rate $F_{\rm haze}$, and 1 bar eddy diffusion coefficients $K_{\rm zz,1bar}$ as free parameters.
For given parameters, the atmospheric temperature-pressure (TP) profile, chemical abundance profiles, and vertical distributions of haze particle sizes and mass mixing ratio are logarithmically interpolated from the grid of photochemical and microphysical calculations using \texttt{RegularGridInterpolator} in the Python scipy library, as utilized in other MANATEE publications \citep{bell2023_methane,welbanks2024wasp107b,beatty24gj3470b,schlawin2024wasp69bpublished}.
We set the nested sampling live points to $400$ and adopt the uniform prior for each model parameter: [M/H]$=\mathcal{U}(2.0,4.0)$, $\log_{10}({\rm C/O})=\mathcal{U}(-1.0,0.05)$, $\log_{10}(F_{\rm haze}~[{\rm g~cm^{-2}~s^{-1}}])=\mathcal{U}(-12.0,-7.0)$, $\log_{10}(K_{\rm zz,1bar}~[{\rm cm^{2}~s^{-1}}])=\mathcal{U}(5.0,9.0)$, and $R_{\rm 10bar}/R_{\rm p,3.6}=\mathcal{U}(0.9,1.0)$, where $R_{\rm p,3.6}=2.7R_{\rm \earth}$ is the approximate transit radius measured at $3.6~{\rm {\mu}m}$ \citep{cloutier2021gj1214bPreciseMass}.
In addition to these planetary parameters, we introduce a constant offset for the transit depth of HST / WFC3, G395H-NRS2 and MIRI-LRS relative to the depth of G395H-NRS1 as other fitting parameters, since there can be an offset among the transit depth measured by each instrument \citep[e.g.,][]{alderson2024compass}.
\rev{Such offset can arise from instrumental effects, such as detector non-linearity that could not be fully removed by polynomial corrections \citep{Schlawin+21_noise2}, as well as the long-term evolution of stellar spot coverage that is indeed reported from the photometric monitoring of GJ1214 \citep{Henry&Bean23,schlawin2024gj1214b_placeholder}.}
We adopt the Gaussian prior centered on zero offset with a standard deviation of $38~{\rm ppm}$ for each instrument offset following \citet{schlawin2024gj1214b_placeholder}.

\section{Column-Integrated Photolysis Rate}\label{Appendix:photolysis}
\begin{figure}
    \centering
    \includegraphics[width=\linewidth]{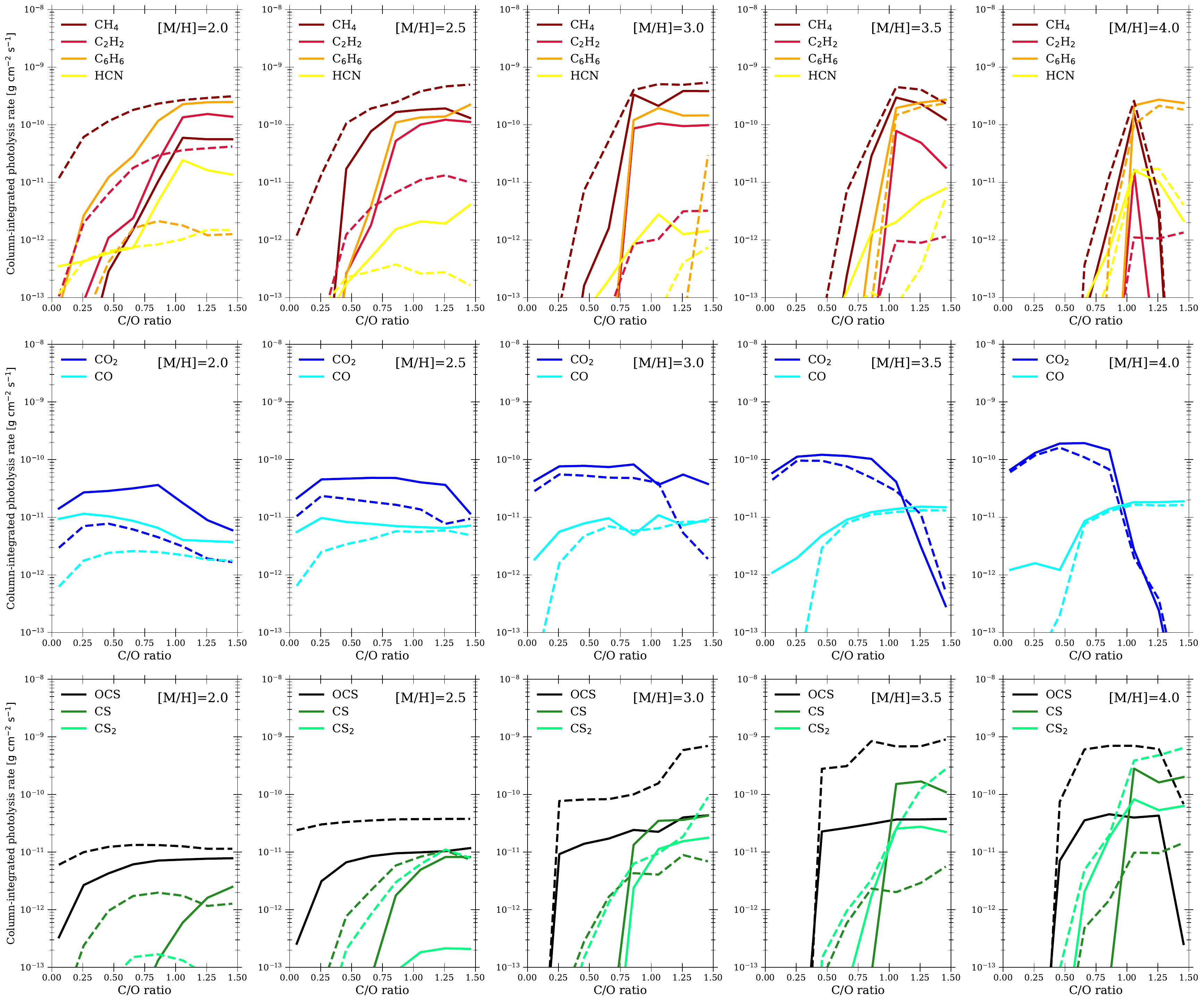}
    \caption{Column-integrated photolysis rates of various carbon-containing species as a function of C/O ratio. From top to bottom, each row shows the photolysis rate of hydrocarbons/nitriles (CH$_4$, C$_2$H$_2$, C$6$H$_6$, HCN), CO and CO$_2$, and sulfur-containing carbons (OCS, CS, CS$_2$). From left to right, each column shows the rates for atmospheric metallicity of [M/H]=2.0, 2.5, 3.0, 3.5, 4.0. The solid and dashed lines show the photolysis rates for the $1~{\rm bar}$ eddy diffusion coefficient of $K_{\rm zz,1bar}={10}^{6}$ and ${10}^{8}~{\rm cm^2~s^{-1}}$, respectively.
    }
    \label{fig:F_spec}
\end{figure}

To assess the mass budget available for haze formation, we follow the approach of previous studies that estimated the possible haze production rate from the column-integrated photolysis rates of hydrocarbons and nitriles \citep{Lavvas&Koskinen17,Kawashima&Ikoma19,Lavvas+19photochemicalHazes}.
This analysis assumes that the fraction of photodissociated molecules end up as solid haze particles.
The column-integrated photolysis rate of each molecule can be calculated by
\begin{equation}
    F_{\rm i}=\int_{0}^{\infty}\frac{m_{\rm i}n_{\rm i}}{\tau_{\rm ph}}Hd\ln{P},
\end{equation}
where $m_{\rm i}$ and $n_{\rm i}$ are the molecular mass and number density for species i, $H$ is the pressure scale height, $\tau_{\rm ph}$ is the photolysis timescale given by
\begin{equation}
    \tau_{\rm ph}^{-1}=\int_{\rm 0}^{\infty}J(\lambda)\sigma_{\rm i}(\lambda)d\lambda,
\end{equation}
where $J(\lambda)$ is the actinc flux density and $\sigma_{\rm i}(\lambda)$ is the molecular photodissociation cross section.

Figure \ref{fig:F_spec} compiles the column-integrated photolysis rates of several key molecules as a function of the metallicity and C/O ratio for $K_{\rm zz,1bar}={10}^{6}$ and $10^8~{\rm cm^2~s^{-1}}$.
In general, the photolysis rates of hydrocarbons and nitriles increases with increasing C/O ratio.
CH$_4$ tends to provide the highest photolysis at lower metallicity with stronger eddy diffusion. 
Although C$_2$H$_2$ was proposed as a leading contributor for hydrocarbon photolysis at metallicity of [M/H]$<2.0$ \citep{Kawashima&Ikoma19,Lavvas+19photochemicalHazes}, we find that C$_6$H$_6$ provide a greater photolysis rate at high metallicity conditions, especially at smaller values of eddy diffusion coefficients.
HCN always has a minor contribution on the hydrocarbon/nitrile photolysis for [M/H]$\ge2.0$, as HCN production requires CH$_4$ and NH$_3$ \citep{Line+11_photochem,moses11}, both of which tend to be depleted at high metallicity atmospheres \citep{Ohno&Fortney23_Nitrogen2}.

The photolysis rates of CO and CO$_2$ are relatively insensitive to C/O ratio and increase with increasing the metallicity.
CO$_2$ photolysis typically dominates over CO photolysis, except for very high C/O ratios for which CO$_2$ is depleted.
CO$_2$ photolysis tends to be lower at stronger eddy diffusion.
This can be attributed to the fact that stronger eddy diffusion more efficiently transport other molecules, such as CH$_4$, to upper atmospheres that act as a UV shield for CO$_2$.

The photolysis rates of sulfur-based species tend to be higher at higher metallicity.
OCS mainly provides a high photolysis rate except for very low C/O and very high C/O at high metallicity environments.
At high metallicity environments, low C/O ratios of C/O$<0.3$--$0.5$ lead to the conversion of sulfur into SO$_2$ rather than OCS, while very high C/O ratio tends to convert OCS into CS$_2$.
CS and CS$_2$ tend to provide high photolysis rates as the C/O ratio increases.
It is interesting to introduce that OCS photolysis is sensitive to $K_{\rm zz}$, and strong eddy diffusion greatly enhances the photolysis rate of OCS.
OCS does not suffer from the UV shielding by other molecules, as the photolysis of OCS takes place at long wavelengths of $>200~{\rm nm}$, which is not overlapped with the photolysis band of hydrocarbons/nitriles and CO/CO$_2$.

\section{Degeneracy due to Potential Instrumental Offsets}\label{Appendix:offset}
\begin{figure}
    \centering
    \includegraphics[width=0.7\linewidth]{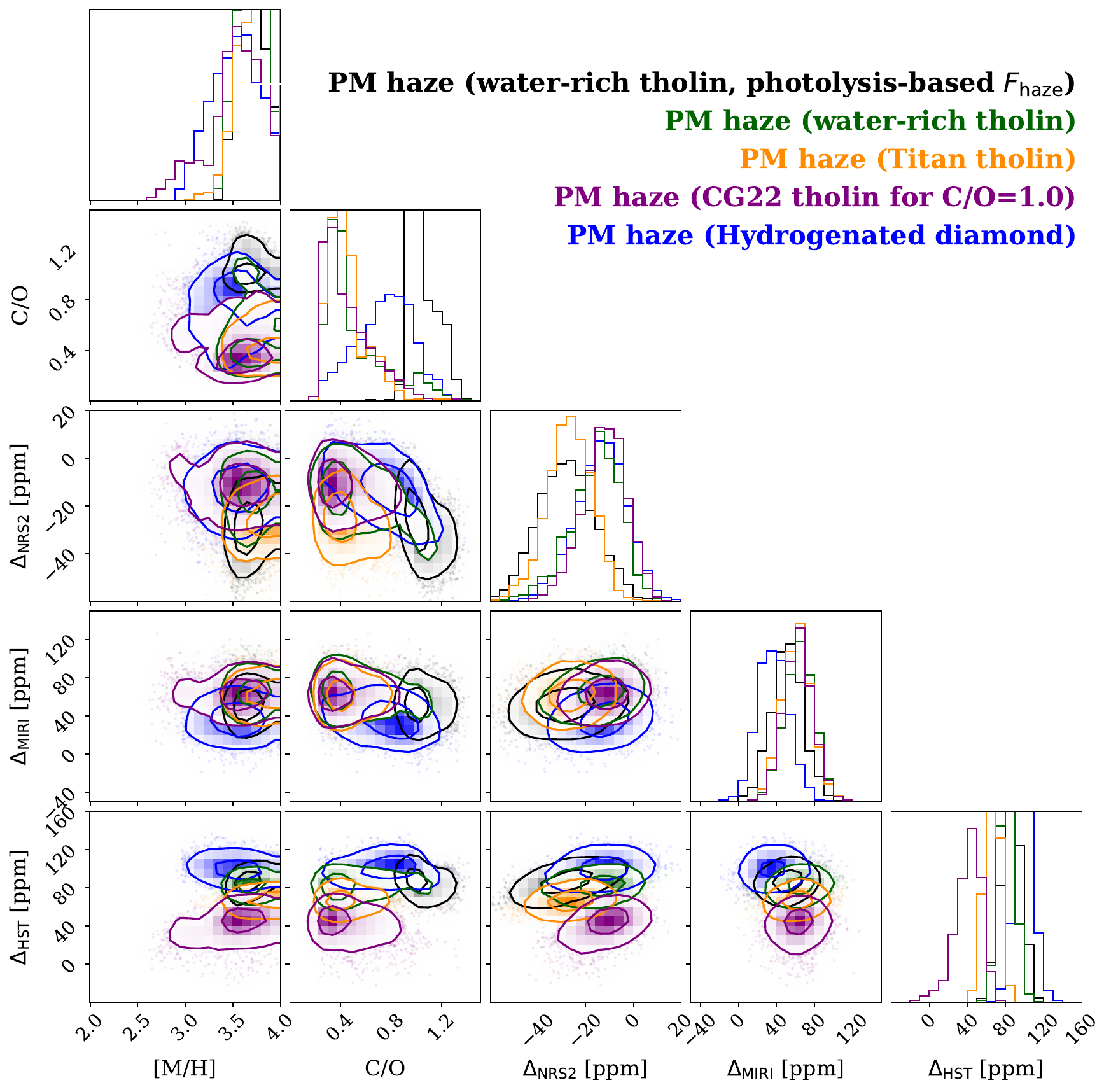}
    \caption{Corner plots for the posteriors of instrumental offsets. The figure include the posteriors obtained for various assumptions of haze production and refractive indices.
    }
    \label{fig:offset}
\end{figure}

We have made potential offsets of HST, MIRI-LRS, and NIRSpec/G395H-NRS2 detector relative to NRS1 detector as free parameters, as these amplitudes remain unknown.
This uncertain instrument offsets limit our ability to constrain haze properties.
Figure \ref{fig:offset} shows the posterior distributions of the instrumental offsets retrieved for different haze refractive indices and treatment of haze production rate.
Although all model configurations consistently suggest the offset of $\Delta_{\rm MIRI}\sim60~{\rm ppm}$ for MIRI-LRS, the required offsets depend on the assumptions of haze properties for NRS2 and HST.
For the detector offset of the NIRSpec/G395H, we obtain the offset of $\Delta_{\rm NRS2}\sim10~{\rm ppm}$ for PM haze grids with water-rich tholin, CG22 tholin, and hydrogenated diamond.
The $\sim10~{\rm ppm}$ offset is consistent with \citet{alderson2024compass} who analysed the potential detector offsets in the NIRSpec/G395H transmission spectrum of super-Earth TOI-836b.
On the other hand, the PH haze grid for Titan tholin and water-rich tholin with photolysis-based haze production requires a larger offset of $\sim30~{\rm ppm}$.
The posterior of HST offset $\Delta_{\rm HST}$ shows diverse values for different assumptions for haze properties, which can be also seen in Figure \ref{fig:trans_many_haze}. 
A better understanding of the instrumental offset would thus help constraining which haze properties are more likely.

We also mention that the uncertain detector offset of NIRSpec/G395H affects the inference of C/O ratio.
For the models with Titan tholin and water-rich tholin with photolysis-based haze production, we can see the correlation between the C/O ratio and the offset between NRS1 and NRS2 detectors in the corner plot of Figure \ref{fig:offset}.
This correlation emerges because CO$_2$ abundance is sensitive to C/O at high metallicity with C/O$\ga$0.5 (see also Figure \ref{fig:VULCAN}), which makes the amplitude of the $4.3~{\rm {\mu}m}$ CO$_2$ feature being dependent of the C/O ratio.
The mismatch due to a smaller CO$_2$ feature at a higher C/O ratio can be somewhat mitigated by a larger offset of the NRS2 detector, which causes the degeneracy between the C/O and the detector offset, especially when the CO$_2$ feature has a small amplitude, say $\la50~{\rm ppm}$, like GJ1214b.
This further highlights the importance of better understanding of instrumental offsets to robustly constrain the atmospheric properties of heavily hazy/cloudy planets.

\end{document}